\documentclass{ws-ijmpa}
\usepackage[super,compress]{cite}
\usepackage{graphicx}
\begin{document}
\markboth{A. De R\'ujula}{QCD, from its inception to its stubbornly unsolved problems}

%
\catchline{}{}{}{}{}
%

\title{\bf QCD, 
from its inception to its stubbornly unsolved problems}

\author{A. DE R\'UJULA
}


\address{Instituto de F\'isica Te\'orica (UAM/CSIC), Universidad Aut\'onoma de Madrid, Spain;\\
Theory Division, CERN, CH 1211 Geneva 23, Switzerland
\\
alvaro.derujula@cern.ch}

%

\maketitle

\begin{history}
\received{Day Month Year}
\revised{Day Month Year}
\end{history}

\begin{abstract}
Whenever one has witnessed some event and then sees it reported in the media,
one's reaction is the same: {\it it was not quite like that.} It is in this spirit of a
frequent first-hand witness that I write this article. 
I discuss a few selected points which
--to my judgement-- illustrate well the QCD evolution (in time) from the 
theoretical, phenomenological and experimental points of view.

\end{abstract}

\ccode{PACS numbers:}


\section{Foreword}
\label{intro}

This is not a review of QCD, of which there are so many that simply trying to select some of them
to be quoted would be quite an endeavor. Instead, this article is a version of a
colloquium at CERN\cite{talk}, written at the request of the publisher.

Adapting to a fashion in novels and films, the text repeatedly
jumps from the past to the future and back. This is not a choice of style, but an
attempt to organize the document by subject matters.

My having witnessed a good fraction of the QCD developments I describe
may make the text vivid, but somewhat self-serving. Yet, my eminent collaborators 
--who do not need any extra praise-- may perhaps appreciate it.

%

\section{The prehistory of quarks}

Back in 1949 Enrico Fermi and Chen Ning ({\it Frank}) Yang published an impressively prescient 
paper \cite{FY}. They posited the notion that some of the few particles then known might not
be elementary, but composite. They assumed that pions are made of a nucleon and an antinucleon.
Treating the $\pi^0$ as a $p\bar p$ bound state they figured out that its (then unknown)
parity ought to be opposite to that of the proton. With the two constituents spinning in the
same direction, they foretold the existence of the $\rho$. They also discussed how the mass 
of the pion may be so much lighter than that of a couple of nucleons if the latter pair were placed
in a very deep square-well potential. Fairly good for a five-page-long paper!

Also in 1949 Jack Steinberger published a calculation  \cite{JS} of the lifetime of the $\pi^0$,
which --besides giving a good estimate-- anticipated the study of ``triangle diagram" 
anomalies in quantum field theory, as well as the predictions of
the rate of gluon fusion in the production of a Higgs boson and the width of the 
$H\to\gamma\gamma$ decay. Not bad for someone trained as a chemist. Soon after his
first steps as a particle theorist, Steinberger moved to the University of California 
at Berkeley, where he became an experimentalist, but continued to study neutral pions.
Fermi and Steinberger are examples of the then much weaker dichotomy between experimentalists
and theorists.

\section{The first non-abelian gauge theory}
\label{sec:GR}

\begin{figure}[b]
\centerline{\includegraphics[width=12cm]{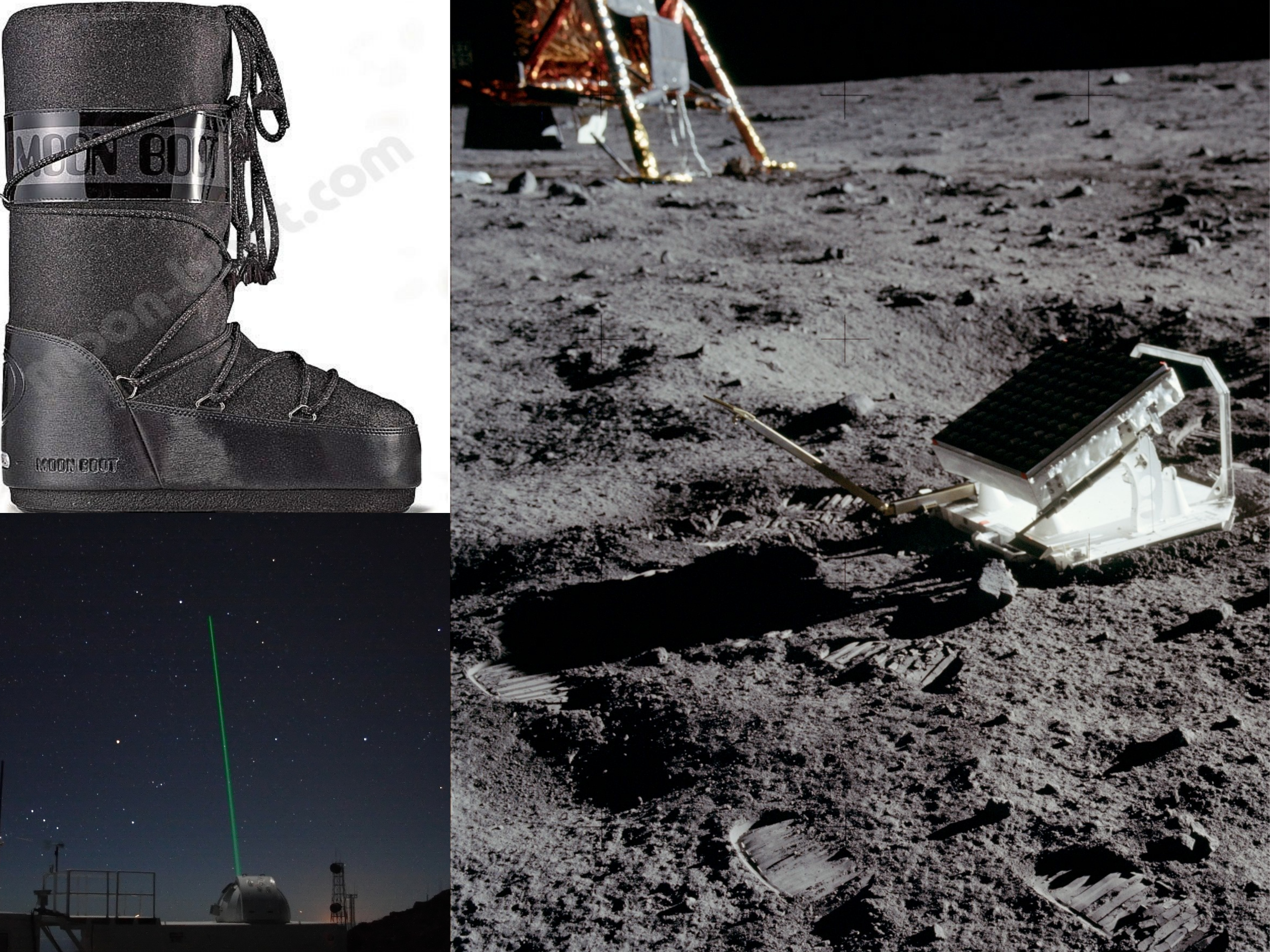}}
\caption{A moonboot, the laser reflector on the Moon, and a laser pointing to it.}
\label{fig-Apolo}     
\end{figure}

Unbeknown to most, the Apollo 11 astronauts actually did something
useful, other than testing Moon boots, as in Fig.~\ref{fig-Apolo}. In 1969, they placed the first 
passive laser reflector on the Moon. After many years of sporadic
lunar ranging measurements, some of the length parameters
describing the lunar orbit are known with millimeter precision.
In a considerable improvement of Galileo's supposed experiment
at the Leaning Tower of Pisa, the 
Earth and the Moon are measured to ``fall'' towards the
Sun with the same acceleration with a precision of $\sim 2\times 10^{-13}$.
This is called the Nordtvedt test of the Equivalence Principle \cite{KN1}.

\begin{figure}[b]
\centerline{\includegraphics[width=12cm]{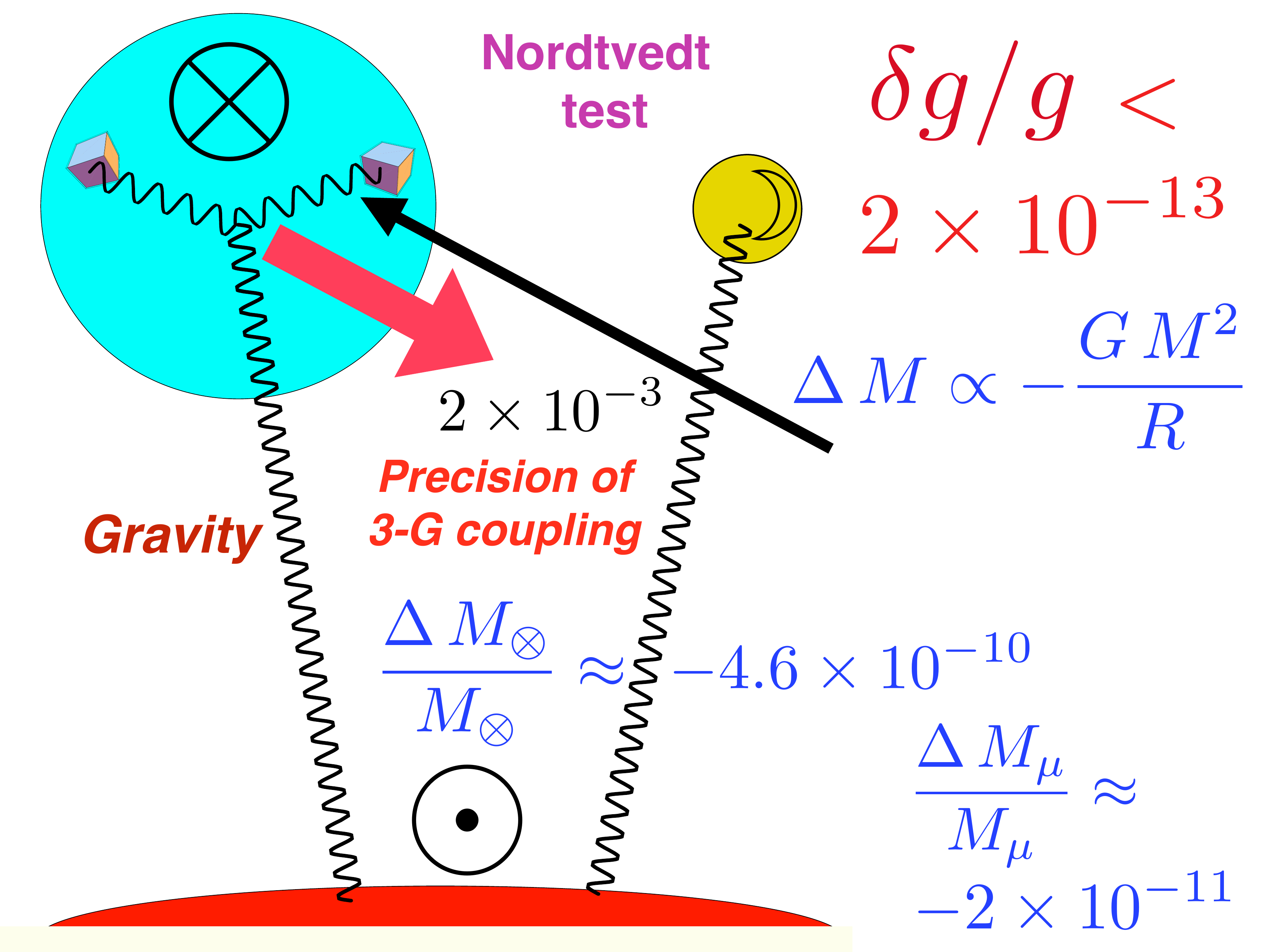}}
\caption{The Earth and the Moon being attracted to the Sun. The triple-graviton
vertex represents the pull by the Sun on the gravitational self-energy
of the Earth.}
\label{fig-TestEP}     
\end{figure}

The gravitational self-mass of a uniform extensive body is 
$\Delta M\propto G_N\, M^2/R$. More precisely, this quantity
is $\Delta M_\otimes\approx -\, 4.6\times 10^{-10}\;M_\otimes$
for our planet, and $\Delta M_\mu\approx -\, 0.2\times 10^{-10}\;M_\mu$
for its satellite. If these self-mass contributions were attracted by the Sun
differently from the bulk of the mass of the two bodies, differing
accelerations would result \cite{KN2}. In actual numbers, we know
that the Earth's bulk acceleration and that of its self-gravity are
equal to a precision of $2\times 10^{-3}$. 

 Einstein's General Relativity (GR) is akin to a
 non-Abelian Yang-Mills theory in the sense that gravitons gravitate.
 The diagrammatic translation \cite{RefA} of the previous paragraph is
 shown in Fig.~\ref{fig-TestEP}. What all this means is that we know the
 triple-graviton coupling to be what it should be, to 2 thousands.
This is better than the precision to which we know the 
``triple-gauge" couplings of intermediate vector bosons or gluons, or
the equality (up to group-theoretical factors) of the colour charges
of quarks and gluons.
There is a long way to go before we have tested Yang-Mills
vector-boson theories to an astronomically satisfactory accuracy!

For the purist I ought to add that in drawing and interpreting Fig.~\ref{fig-TestEP}
I seem to have assumed that gravitons exist (no reason to doubt it). What the Nordtvedt
test really tests is that the energy-momentum tensor of the Coulomb-like gravitational 
self-energies of the Earth and the Moon couple to an external gravitational field with the
strength predicted by GR for this non-abelian coupling. Much as one would derive the 
Coulomb potential from the Fourier transform of a
Feynman diagram involving a known-to-exist photon... I have drawn
and interpreted  Fig.~\ref{fig-TestEP}.

Neil Armstrong and Buzz Aldrin, the astronauts of Apollo 11 who set foot on the Moon, also left a seismometer on its surface. The lunar seismometer, measuring moon-quakes, 
inaugurated the study of the inner properties of celestial bodies other than the Earth. 
Later Apollo crews brought with them more precise instruments of these and other types.

One moral to be extracted from the above is that once upon a time fundamental physics --such as testing the Equivalence Principle-- was considered sufficiently important to include a fairly massive light reflector on the first Moon landing. No need to emphasize how difficult and significant such a ``sacrifice" of weight-to-be-lifted must have been.

\section{Who invented quarks?}

A tricky question. The official history is that they were invented
by Gell-Mann and  Zweig, in that chronological order:
Gell-Mann's published paper \cite{MGM} was received by Physics Letters on January
4th 1964, while Zweig's unpublished work is a CERN yellow report \cite{GZ} dated
January 17th of the same year. But, as Napoleon is said to have said:
{\it History is the version of past events that people have decided to agree
upon.}

According to the same un-trustable source of the previous quote 
(Internet) Gell-Mann's paper was originally rejected by Physical Review Letters. Allegedly
untrue \cite{MGM1}.  None of this would significantly change the chronological order, 
which is anyway fairly irrelevant since the dates were so close. Concerning dates, it must be recalled 
that Gell-Mann wrote: {\it These ideas were developed ... in March 1963; the 
author would like to thank Professor Robert Serber for stimulating them.}
For Serber's recollections, see Ref.~(\refcite{Serber}).

According to Serber's autobiography\cite{Serber}, he knew how to build representations of SU(2)
from the fundamental spinor one. In March 1963 he was trying to extend this
construction for SU(3) and he mentioned all this to Gell-Mann, who promptly realized
that the components of the required 3 and $\bar 3$ representations of SU(3) would have 
to have fractional charges. Their existence 
{\it would be a strange quirk of nature, and
quirk was jokingly transformed into quark.}\cite{Serber}

A point in the official history  \cite{Off} is lacking \cite{RefA,CC}.  Andr\'e Petermann
published a paper (in French!) \cite{Petermann},
received December 30th, 1963, shortly before the dates quoted
in the first paragraph of this section.
In this paper he discussed mesons as made of a {\it spinor/anti-spinor
pair} and baryons as {\it composed of at least three spinors.} Concerning
the delicate issue of their charges, Petermann delightfully writes:
{\it if one wants to preserve charge
conservation, which is highly desirable, the spinors must have
fractional charges. This fact is unpleasant, but cannot, after all, be
excluded on physical grounds.}

There are other unofficial issues concerning this chapter 
of the history of science. Was Zweig forbidden to have a preprint typed and to give a talk
at CERN at the time? If so, by whom? I shall not answer these questions,
but another one which I have been challenged to answer:
why was the publication of Petermann's paper delayed for a year?
Alas, nobody has found the original CERN preprint, yet.
So, one cannot disprove something evil I have been told, namely that {\it Petermann 
had plenty of time to change the paper before publication.}
That was not his style. Not only did he publish this paper in French
--guaranteeing a dearth of readers-- but, though he always drove a Porsche
and on occasion coached the Swiss ski team, he was
extremely slow at anything else, including bothering to correct the proofs of an 
article\footnote{Having been a coauthor of Peterman's, I know this first hand. Notice
also that I have spelled his name differently here. He did not even care to sign all
his papers with the same name.}.

\section{Searching for free quarks}

Gell-Mann upheld, perhaps for longer than anybody else, the view  that quarks
were mere mathematical objects. After all, the notion that hadrons are made of parts 
--but cannot be taken 
apart-- is not that easy to accept, even grammatically. But none of this deterred
experimentalists from looking for unconfined quarks and dozens of searches were
performed \cite{pdgQuarks}.

 Amongst the many experiments hunting for free quarks,
 perhaps the most original one was based on analyzing oysters; it was conducted
 by Peter Franken and collaborators \cite{PFQuaks}.
 The liver of an oyster, it is stated, is one of the best filters
 anywhere. Every day it processes an amount of water some
 thousand times its weight. In so doing, it accumulates
 all sorts of peculiar substances that are preserved in its growing shell. An atom or molecule
 containing one extra quark would have a fractional electrical
 charge and should have a very peculiar
 chemistry. It would presumably be sieved by the oyster's liver.
 Alas, these experimentalists did not find any quarks. But, in the
 process of studying a barrel of New England oysters a day, they probably
 gained some weight.
 
Not all experiments failed in finding evidence for fractionally charged objects. Perhaps
the most notorious ``successful" search was the one performed 
by William Fairbank and collaborators \cite{Fquarks}.
It employed the venerable Millikan technique, using Niobium-coated Tungsten balls,
and found one with charge $(0.337\pm 0.009)$e. Needless to say (now) the finding could not be
reproduced.

\section{Other quark mysteries, also unveiled}

The naive (constituent) quark model was impressively successful in its understanding of hadrons made of
$u$, $d$ and $s$ quarks and in predicting the existence, decay pattern and mass of the $\Omega^-$ \cite{M&Y}. But quarks and their confinement remained mysterious, more so because of the complementary evidence in SLAC's deep-inelastic electron-scattering experiments for charged constituents
of protons \cite{Dis}, Feynman's {\it partons} \cite{RF}, with ``point-like"  
interactions with photons: Bj{\o}rken's {\it scaling} \cite{BjF,BjF2}.

Since it has been done so very many times,
I shall not discuss the original literature on Yang-Mills theories \cite{YM}, QCD 
\cite{G1,G2,G3},
the electro-weak standard model \cite{SLG,Steve,AS,AS2},
the necessary existence of charmed quarks \cite{GIM}
and the renormalizability of non-abelian gauge theories \cite{'t,'tV}. 

The discovery of 
strangeness-conserving neutral currents in neutrino scattering
by the Gargamelle bubble-chamber collaboration at CERN \cite{Musset} 
made experimentalists, and the world at large, aware of Yang-Mills theories, 
much as the 1971 work of 't Hooft \cite{'t} and 't Hooft and Veltman \cite{'tV} 
immediately attracted attention
from (field) theorists to the same subject. For the hypothetical young reader I must emphasize 
that, at the time, the fact that the Standard Model had all the chances of being ``right"
was only obvious to an overwhelmed minority of field-theory addicts.

The understanding
of how quarks behaved when probed at short distances had to wait for the 
{\bf discovery}\footnote{The fact that QCD's asymptotic freedom 
was first noticed by 't Hooft, Symanzik and perhaps others made me emphasize ``discovery"
which, when not italicized, includes the realization of how important something may be.}
 of QCD's asymptotic freedom \cite{DDF1,DDF2}. At the time 
 David Politzer's office was next to mine at Harvard. 
 David Gross and Frank Wilczek were at Princeton. The Harvard/Princeton competition
 was acute \cite{DPNL} and productive \cite{RefA}. To characterize it,
 suffice it to say that Harvard's
 motto is {\it VERITAS} (Truth), while Princeton's is {\it DEI SVB NVMINE VIGET} (God went to Princeton).
 
In a talk reproduced in Ref.~\refcite{HGeorgi}, Howard Georgi
recalled how everybody, in years long past, 
knew Harvard as the place {\it not} to be. He was the seventh of a long list
of applicants. The first six had chosen ``better'' destinations. One year later,
I was to share Howard's honour. It turns out that I was not quite at the
right place at the right time but I was, literally, next door.  Indeed, when
QCD's asymptotic freedom was discovered David Politzer 
had the office next to mine at Harvard. He was a Junior Fellow and I a lowly
post-doc. Some of the  outcasts that gathered in this back-door way changed 
physics (and Harvard) forever. Now Harvard is again {\it a place
where to be}, but for far far more formal reasons.

In the late '60s, it seemed perfectly ridiculous for the strongly interacting partonic
constituents of protons to do what they do: exhibit a ``scaling''
free-field behaviour  \cite{BjF,BjF2} in deep inelastic scattering experiments
 \cite{Dis}. Thus, though the full rationale for a rather low-energy
``asymptotia'' remained obscure for a while, the discovery of
asymptotic freedom was received with a great sigh of relief by field-theorists.

\section{A renewed call for leniency }
\label{sec-1}

Hereafter I am going to cite papers in an unbalanced way, with a large fraction of references to articles authored or coauthored by me. Part of this is a proximity effect, I am writing as a witness and a $1/r^2$ law is inevitable, a price to pay for personal recollections, often more colorful than ``official" histories.

As Golda Meyer put it: {\it Don't be humble... you are not that great}. That is correct in my case, but it does 
not apply to any of my to-be-cited coauthors, as the reader will easily recognize.

\section{ $\mathbf {\alpha_s}$ and $\mathbf{\Lambda}_{\rm\bf QCD}$}

The first concrete predictions of QCD  \cite{moments,moments1,moments2,moments3}
concerned the deviations from an exact scaling behaviour. But the electron
scattering and $e^+e^-$ annihilation data of the time  \cite{Dis} covered
momentum transfers, $Q^2$, of not more  than a few GeV$^2$. Nobody (yet) 
dared
to analyze these data in the ``asymptotic'' spirit of QCD. And that is how
some people not affected by dataphobia ---a morbid condition of the brain
(or brane?) that turns theoretical physicists into mathematicians--- set  out to
exploit the only data then available at higher $Q^2$.

\begin{figure}[b]
\centerline{\includegraphics[width=12cm]{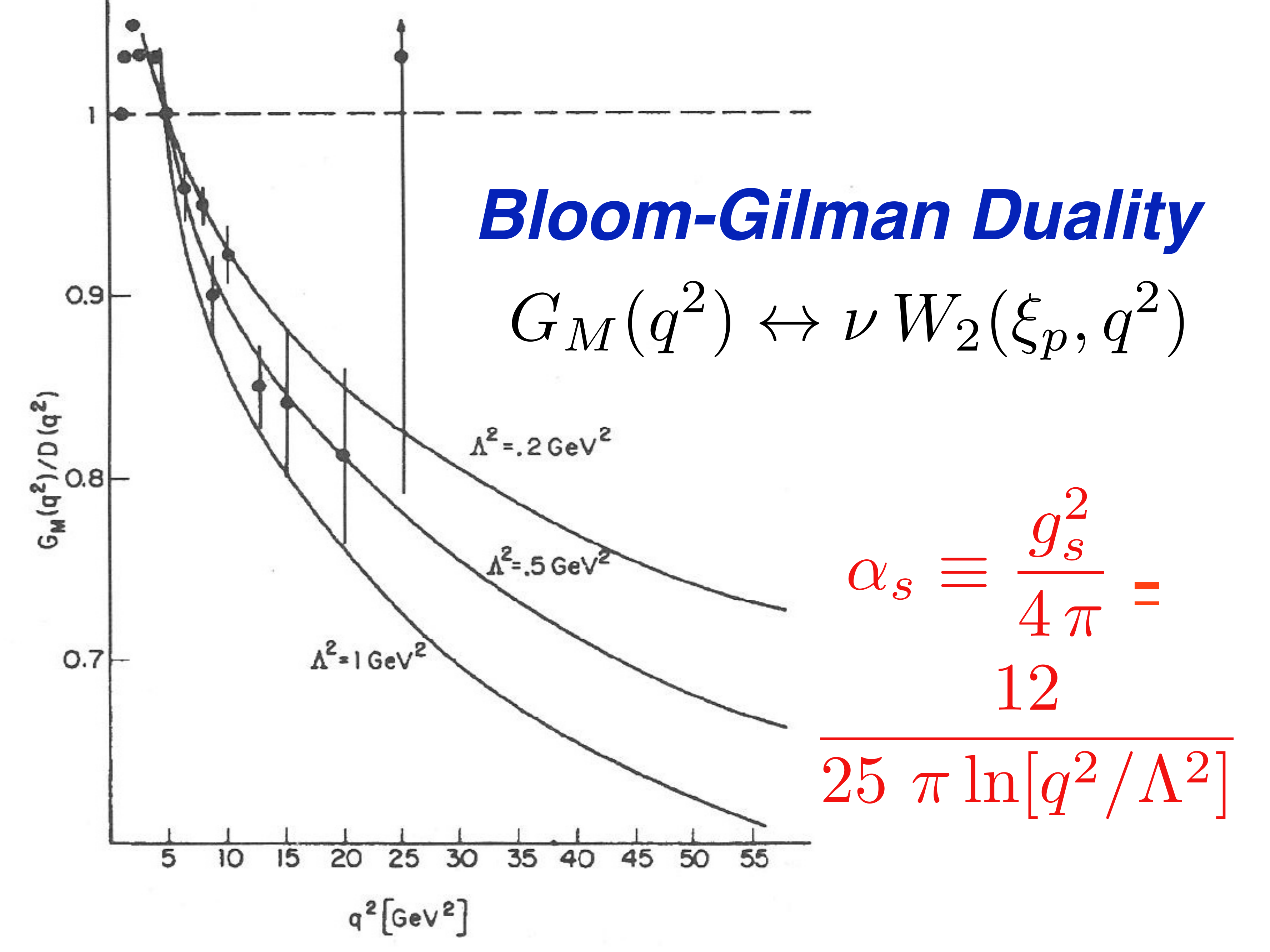}}
\caption{The proton form factor $G_M(q^2)$, divided by the usual dipole parametrization
$D(q^2)$. Un-normalized results for $G_M(q^2)$ would show the agreement between
theory and data over a much more extended range on the vertical axis.}
\label{fig-FF}     
\end{figure}

By the early '70s, the proton's elastic form factor had been measured
 \cite{proton} up to $Q^2\sim 20$ GeV$^2$. To bridge the gap between the QCD
predictions for deep inelastic scattering and the elastic form factor, two
groups  \cite{yo,GT} used (or, with the benefit of hindsight, slightly abused) the
then-mysterious ``Bloom--Gilman duality''  \cite{BG,BG1} relating the deep
``scaling'' data to the elastic and quasi-elastic peaks. I prefer the paper
containing Fig.~\ref{fig-FF} and beginning:  {\it ``Two virtues of asymptotically 
free gauge theories of the strong interactions are that they are {\bf not}
free-field theories and they make predictions that are {\bf not}
asymptotic''}\footnote{The fact that one felt obliged to emphasize all this means that it was
not at all obvious to the community at the time.}; to conclude {\it ``The results agree with experiment but are
{\bf not} a conclusive test of asymptotic freedom.''} 

Not atypical of the Harvard/Princeton
 competition of those times, the papers I just quoted
 \cite{yo,GT} were received by the publisher within a one-day interval
(mine was the late one). These are the first two papers on QCD phenomenology. They
were written a full year after the discovery of asymptotic freedom. It is quite an
amazing coincidence that they were so well  synchronized.

Being a bit more inclined to data analysis than my Princeton competitors
--and wanting to be the first theoretical physicist  [ (-: ]  
to extract from observations a fundamental constant of nature--
I obtained a value and an error range for 
$\Lambda\equiv\Lambda_{\rm QCD}$,  while David Gross and Sam Treiman
simply chose a reference value for this quantity (which they called $\mu$),
perhaps because their results ---based on an analysis slightly different from mine---
neither fitted the data nor subtracted from their confidence in the theory
\cite{GT}. 

\begin{figure}[b]
\centerline{\includegraphics[width=12cm]{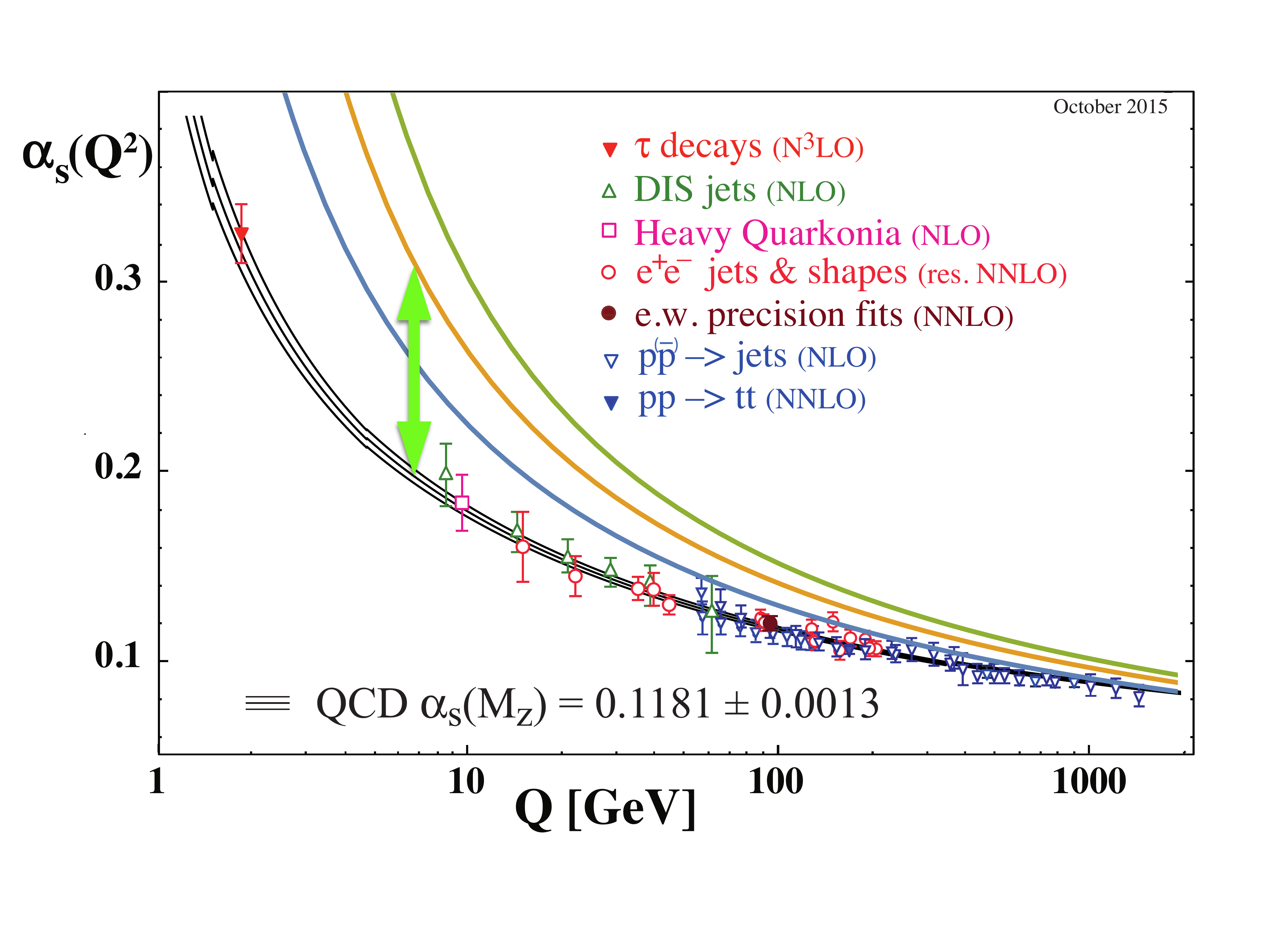}}
\caption{Recent\cite{alphas} (black lines) and early\cite{yo} (colored lines) results on $\alpha_s(Q^2)$. }
\label{fig-Lambda}     
\end{figure}

In Fig.~\ref{fig-Lambda} I show recent and very sophisticated results \cite{alphas}
on QCD's fine-structure 
``constant", $\alpha_s(Q^2)$, as well as my original three-flavored leading-order result, 
$\alpha_s=12/[25\,\pi\log(Q^2/ \Lambda^2)]$, and its error range. The vertical green double
arrow shows, that the central value of the original determination 
of $\alpha_s$ must be reduced by $\sim 33\%$ to get close to best current results.
This level of discrepancy between leading-log QCD results and more sophisticated
ones is quite typical. 

One may also recall that dimensional transmutation \cite{Sidney} was also very hard
for many to accept. It is the ab-initio astonishing renormalization-group 
fact that, in an asymptotically free
theory, a dimensionless coupling defined by its value at a given momentum scale 
(two parameters) is equivalent to a one-parameter
expression at different momentum scales: $\alpha_s(Q^2/\Lambda^2)$ as quoted above,
 in this leading-log case.

\section{Bloom--Gilman duality (BGD)}

Is Bloom--Gilman duality \cite{BG,BG1} a prediction of QCD? In spite of recent 
efforts \cite{HCM}, it is not (yet).
That would require a complete understanding of bound-state production. But 
perturbative QCD {\it explains} BGD, in its QCD-improved realization
\cite{DGP2}. If you trust me, do not
read this technical section, but for the rest of this paragraph.
Two of the results of Ref.~\refcite{DGP2} are worth recalling. The first is that our analysis
of deep inelastic electron scattering resulted in $\Lambda=500\pm 200$ MeV,
in agreement with the results of Ref.~\refcite{yo} and Fig.~\ref{fig-Lambda}. 
The second is that we also extracted
results to Next-to-Leading-Twist (NLT). These are characterized by a mass $M_0$ that ought
to be of ${\cal{O}}(\Lambda)$. We obtained $M_0=375\pm 25$ MeV.
As data on other processes become more precise, a revival of NLT analyses may become mandatory.

\begin{figure}[b]
\centerline{\hspace{4cm}\includegraphics[width=14cm]{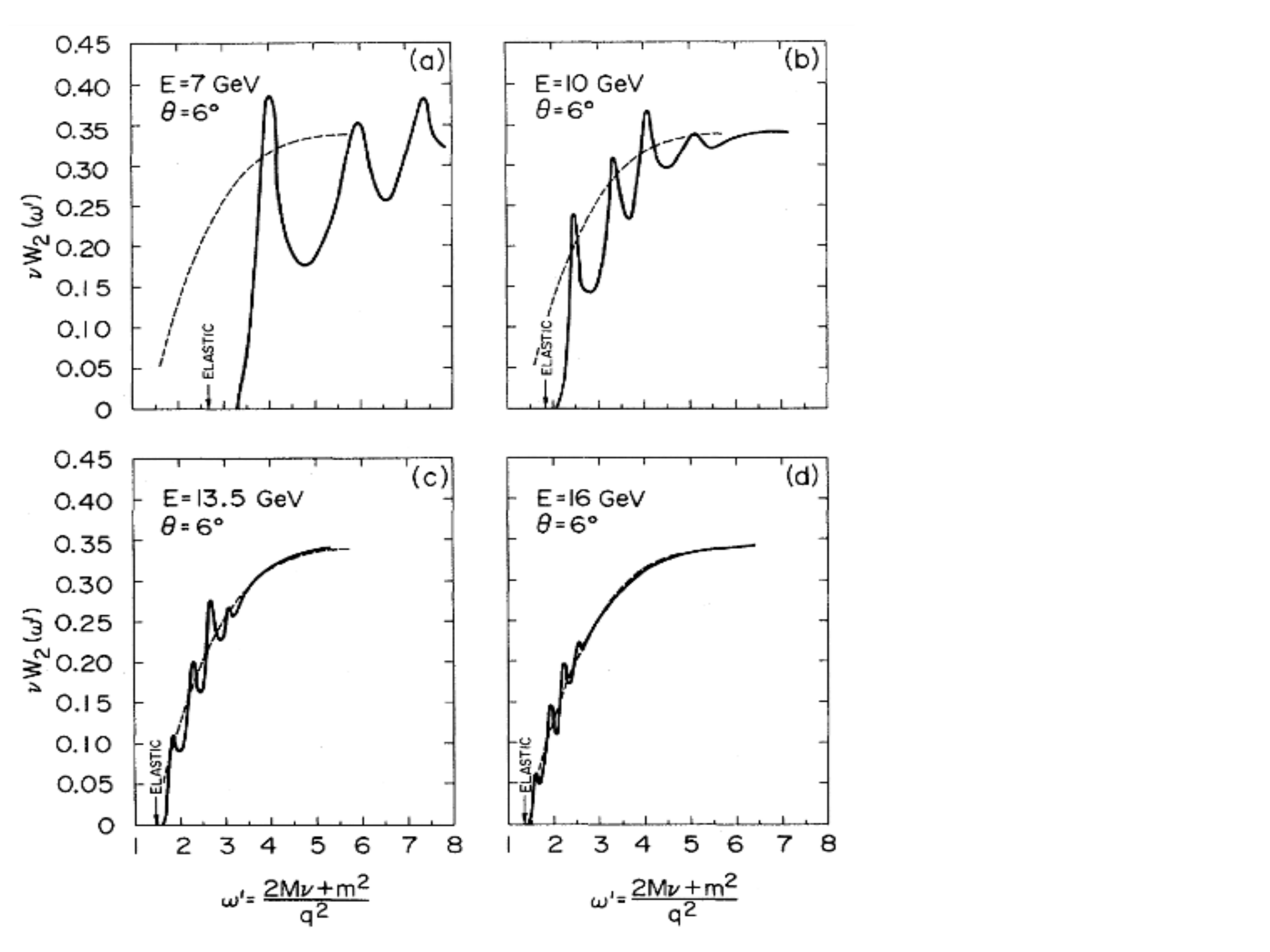}}
\caption{Bloom-Gilman duality\cite{BG}  at a fixed $e$-scattering angle and varying energies
(or $Q^2$ values).}
\label{fig-BG}     
\end{figure}

BGD is the observation \cite{BG,BG1} that at low $Q^2$ a structure function
shows prominent nucleon resonances, which ``average'' to the
``scaling'' function measured at some higher $Q_0^2$, and snuggly slide down
its slope as $Q^2$ increases. 
As shown in Fig.~\ref{fig-BG}, this happens if the chosen scaling variable
in not Bj{\o}rken's $x=Q^2/(2\,m_p\,\nu)$, but contains a ``target mass correction",
$1/\omega' \equiv x'=Q^2/(2\,m_p\,\nu+m_p^2)$. All this was considered at the time,
in Californian style, as mystifying as a myth.

In  {\it Demythification of Electroproduction Local Duality and Precocious Scaling}
\cite{DGP2} Howard Georgi, Politzer and I argued that BGD is a consequence of QCD,
inevitable if scaling is ``precocious'', as it must be for small
$\Lambda$ (a fraction of a GeV). 

\begin{figure}[b]
\centerline{\hspace{.4cm}\includegraphics[width=14cm]{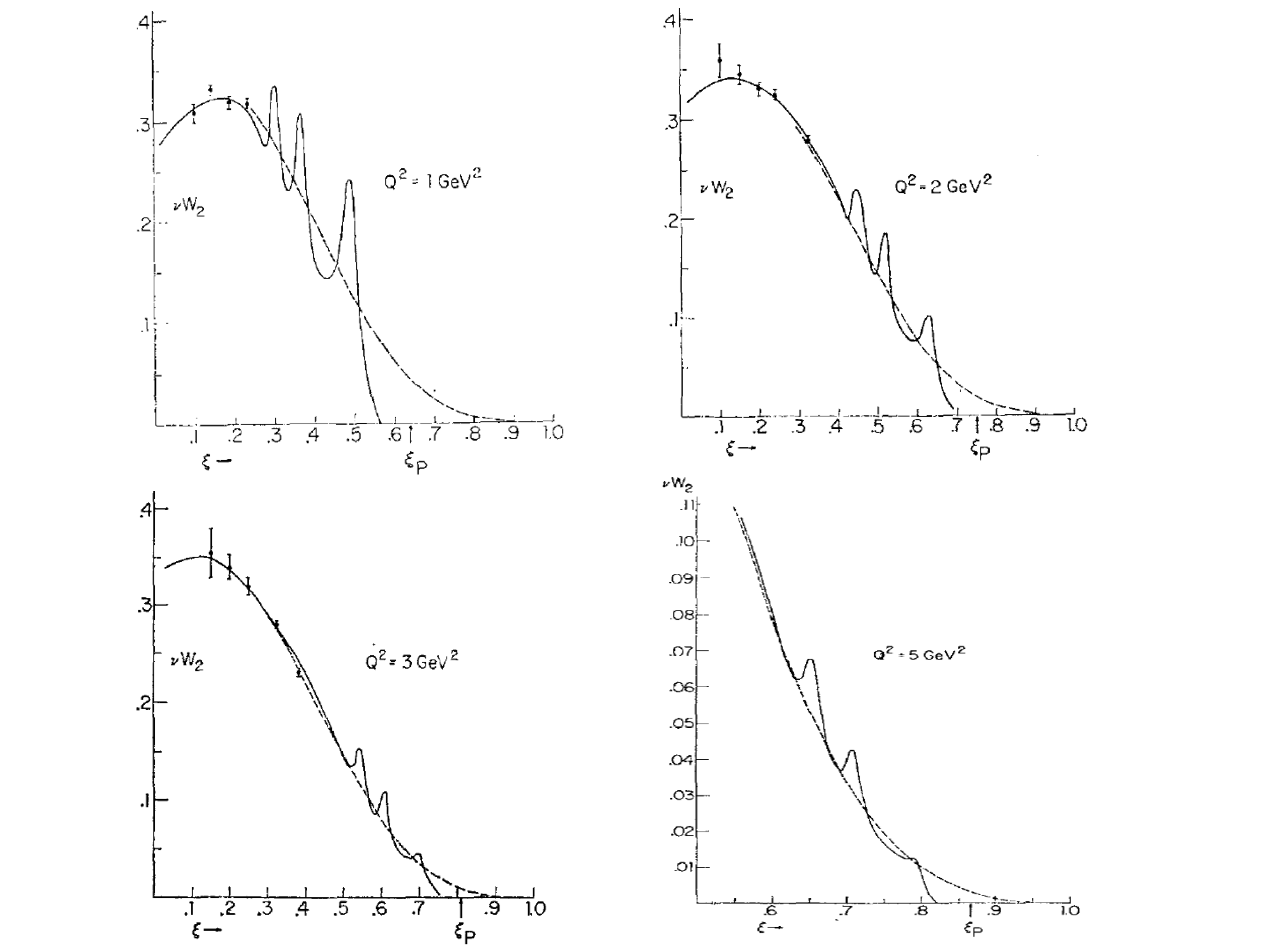}}
\caption{Dashed line: the perturbatively evolved\cite{DGP2} proton structure function $\nu W_2(\xi,Q^2)$.
Continuous line: a fit to actual data. $\xi_p$ is the position of the elastic 
contribution $\propto G_{\!M}^2\delta(\xi-\xi_p)$. A few data points are also shown.}
\label{fig-BGD}     
\end{figure}

The scaling variable to be used, as we insisted in Ref.~(\refcite{xi}), 
is not $x'$, but the one implied by a full use of QCD's operator-product expansion, i.e.~Nachtman's 
variable \cite{Otto} $\xi= 2\,x/[1+(1+4\,m_p^2\,x^2/Q^2)^{1/2}]$, which takes care of the target--mass
``higher-twist'' effects of order $m_p^2/Q^2$. In studying the $Q^2$-evolution of 
the $n$-th moment of  a structure function, weights $\xi^n$ and not $x^n$ ought to be used.
And the entire structure function is to be $\xi^n$-weighed, including the elastic contribution at 
$\xi=\xi_p\equiv 2/[1+(1+4\,m_p^2/Q^2)^{1/2}]$. The QCD duality is shown in Fig.~\ref{fig-BGD}.

The crucial point is that the customary logarithmic QCD evolution of structure functions
 has higher-twist corrections. The next to leading-twist ones are of the form 
$(1 + n\,a_n\, \Lambda^2/Q^2)$, with $|a_n|\simeq 1$, as the data allowed us to 
check in Ref.~[\refcite{DGP2}].
Consider taking the $n$-th $\xi$-moment of a structure function measured at a 
relatively large $Q_0^2$, where the resonant peaks are barely observable.
Next, evolve this moment perturbatively down to a lower $Q^2\ll Q_0^2$.
Since the $a_n$ are not perturbatively calculable, the perturbative prediction for the $n$th moment
at the scale $Q^2$ has a relative uncertainty of ${\cal O}(n\,\Lambda^2/Q^2)$,
where the factor $n$ is crucial.

\begin{figure}[b]
\centerline{\includegraphics[width=12cm]{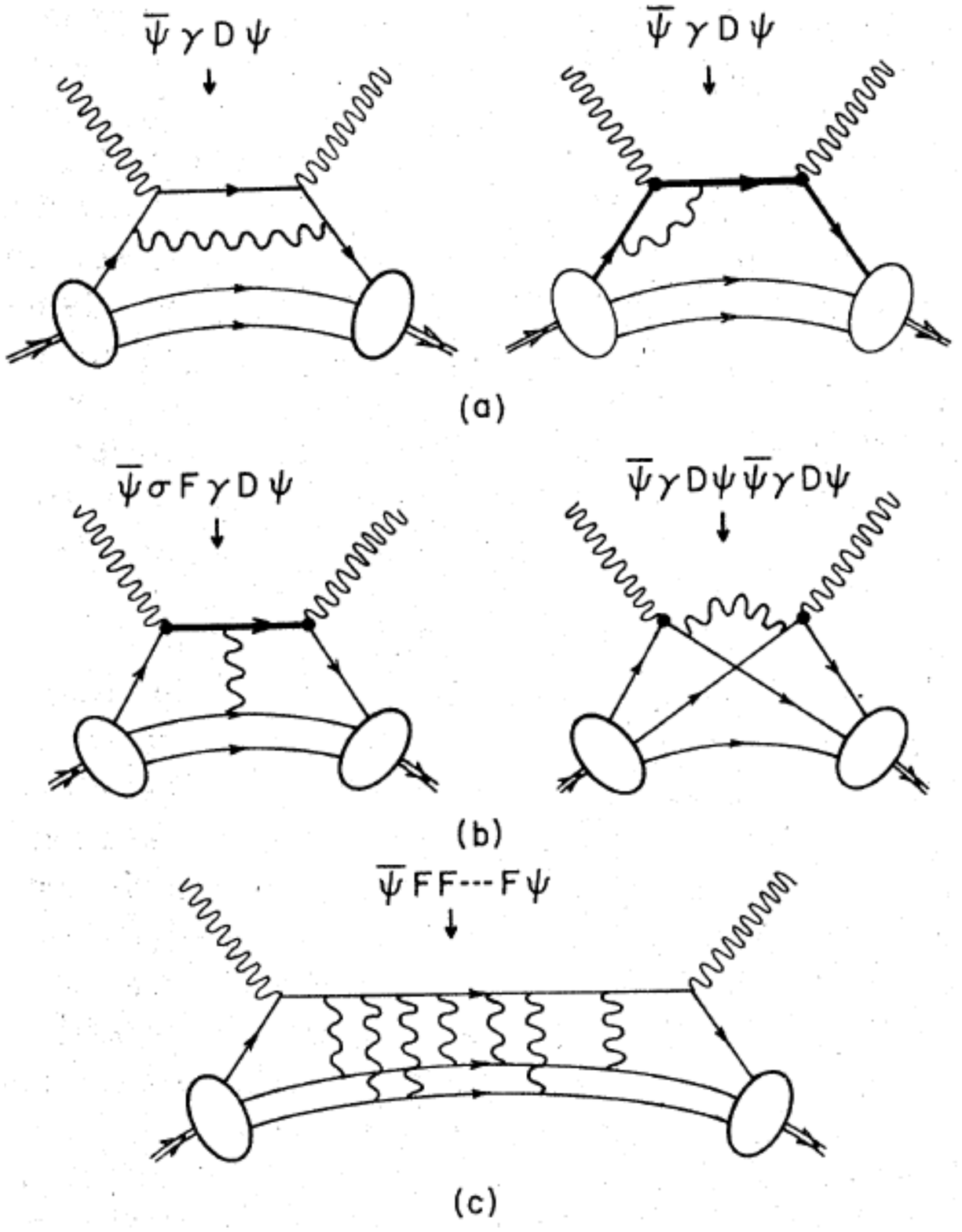}}
\vspace{-4cm}
\caption{Operators and parton-language diagrams at various twists\cite{xiright}.
(a) shows typical twist-2 effects, corresponding
to a parton-model picture with no communication between
struck quarks and spectator quarks. (b) shows
twist-4 effects. (c) shows an even  higher-twist effect.}
\label{fig-HT}     
\end{figure}

Our claim\cite{DGP2} that Bloom-Gilman duality and precocious scaling were consequences
of QCD must have looked too good to be true, for it
met with an immediate barrage of theoretical papers attempting to prove us 
wrong\cite{EPP,BEGR,GTW}. In Ref.~\refcite{xiright} we defused all this artillery and,
more constructively, we used the parton-model language to interpret the field-theoretic
operator-product expansion in successive twists while developing an intuitive physical 
interpretation of twist, illustrated in Fig.~\ref{fig-HT}. We did also reanalyze the 
``paradoxes" in Ref.~\refcite{GTW} in parton language and detailed how to deal with
the production of hadrons containing quarks heavier than the proton. Finally, we resolved the 
question of nonperturbative effects in the analysis of electroproduction.

\begin{figure}[b]
\centerline{\hspace{1.cm}\includegraphics[width=8cm]{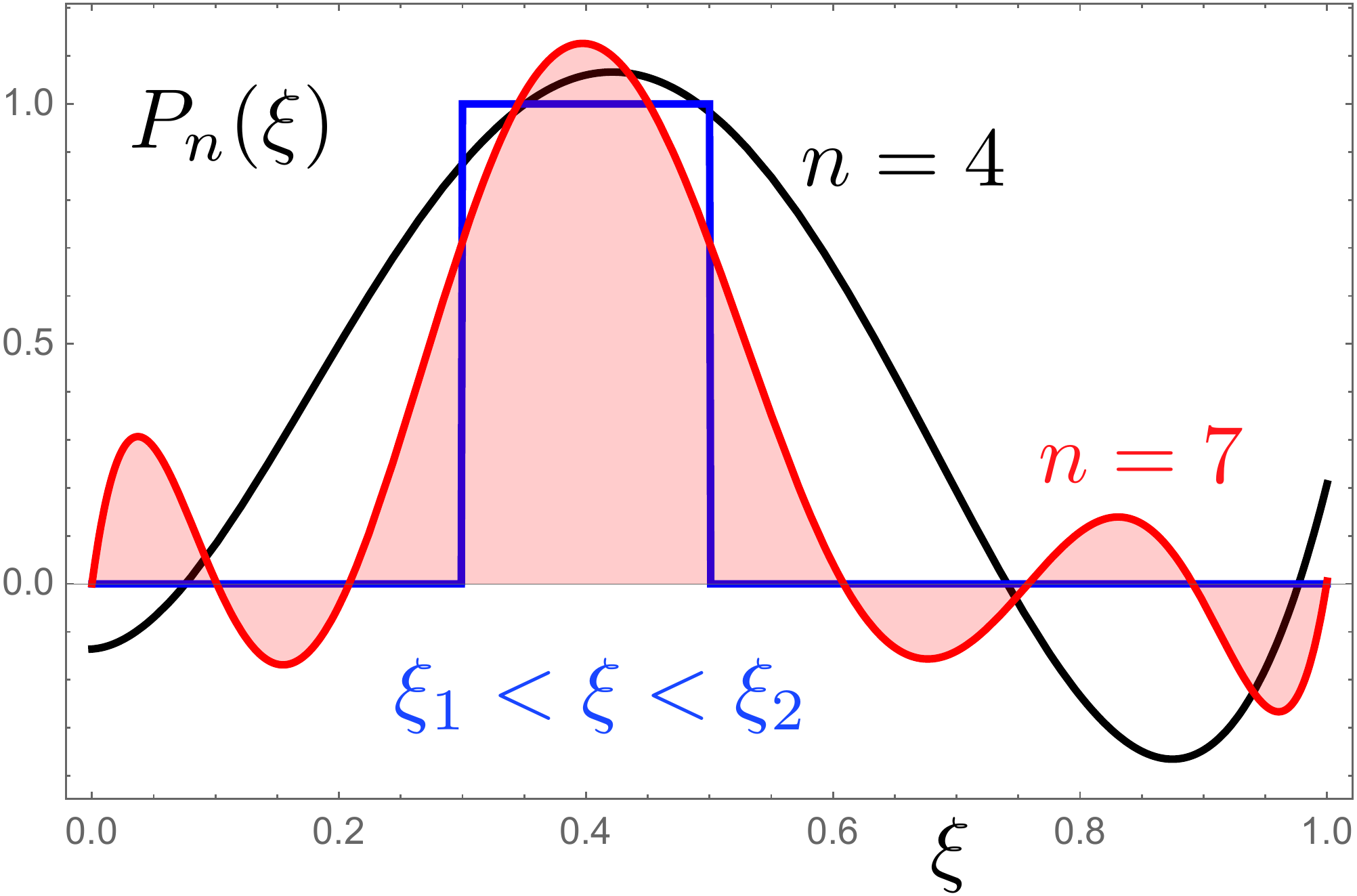}}
\caption{Best fits to a window function from $\xi_1\,{\rm to}\; \xi_2$
with polynomials $P_n(\xi)$, for $n=4$ and 7.}
\label{fig-Polynomial}     
\end{figure}

To extract information on local duality from $n$ available moments, consider
 a polynomial $P_n(\xi)= \sum_0^n C_m\,\xi^m$.
One can find, for a given $n$, the $C_m$'s corresponding to a best fit to a ``window function"
that can be used to test duality ``locally'', i.e.~in a chosen interval $\xi_1\,{\rm to}\; \xi_2$, 
see Fig.~\ref{fig-Polynomial}.
Given a predicted set of moments with uncertainties of ${\cal O}(n\,\Lambda^2/Q^2)$
one expects  a more local and precise QCD duality 
the smaller $n/Q^2$ is. That is precisely what is observed \cite{DGP2}. QED\footnote{
The actual analysis of QCD duality is a bit more elaborate, since one expects
slightly different precisions for window functions centered at different $\xi$'s \cite{DGP2}.}.

The preceding detailed discussion justifies a posteriori
the analysis of Ref.~\refcite{yo}, based on BG local duality in an interval
enclosing the elastic proton contribution $\propto G_{\!M}^2 \delta(\xi-\xi_p)$.
It also explains why these initial attempts 
at QCD phenomenology resulted in reasonable and consistent values of $\Lambda$.

\section{Progress on the determination of $\mathbf {\alpha_s}$}
\label{sec:alphas}

In this section I rely heavily on the contribution by Bethke, Disertori and Salam
in the Review of Particle Properties \cite{BDS}. 

As a function of a scale $\mu_R$,
the renormalization group improved perturbation expansion of $\alpha_s$ reads:
\begin{equation}
\mu_{ R}^2{d\alpha_s\over d\mu_{ R}^2}=
\beta(\alpha_s)=-(b_0\,\alpha_s^2+b_1\,\alpha_s^3+b_2\,\alpha_s^4+b_3\,\alpha_s^5+...),
\label{eq:dalpha}
\end{equation}
where all the quoted $b_i$ have been calculated and the minus sign is of asymptotically
free fame ($b_0>0$ for fewer than 17 flavors).

The quantity $\alpha_s(Q^2)$, measured at a specified momentum scale is measurable.
In a limited sense that is, for the coefficients $b_i$, $i{\geq 2}$ in Eq.~(\ref{eq:dalpha}) are 
scheme-dependent, a first sign of discomfort. One of the most precise measurements of
$\alpha_s(Q^2)$ relies on the illustrious ratio $R$ and the precise knowledge of $R_{EW}$,
its value for a free-quark ansatz with $\alpha_s=0$:
\begin{equation}
{\sigma(e^+e^-\to{\rm hadrons},Q) \over
\sigma(e^+e^-\to \mu^+\mu^-,Q)} \equiv R(Q)
=R_{\rm EW}(Q)[1+\delta_{\rm QCD}(Q)].
\label{eq:R}
\end{equation}
The pertubative series for $\delta_{\rm QCD}(Q)$ is:
\begin{equation}
\delta_{\rm QCD}(Q)=\sum_{n=1}^\infty
c_n \, \left[{\alpha_s(Q^2)\over \pi}\right]^n
+{\cal{O}}\left[{\Lambda^4\over Q^4}\right]
\label{eq:delta}
\end{equation}
The coefficients $c_1$ to $c_4$ have been calculated \cite{Baikov}.
The series converges slowly, perhaps due to ``renormalon" divergences,
related to non-perturbative contributions and behaving as $n!\,\alpha_s^n$,
another current limitation \cite{Beneke}.

\begin{figure}[b]
\centerline{
\includegraphics[width=11cm]{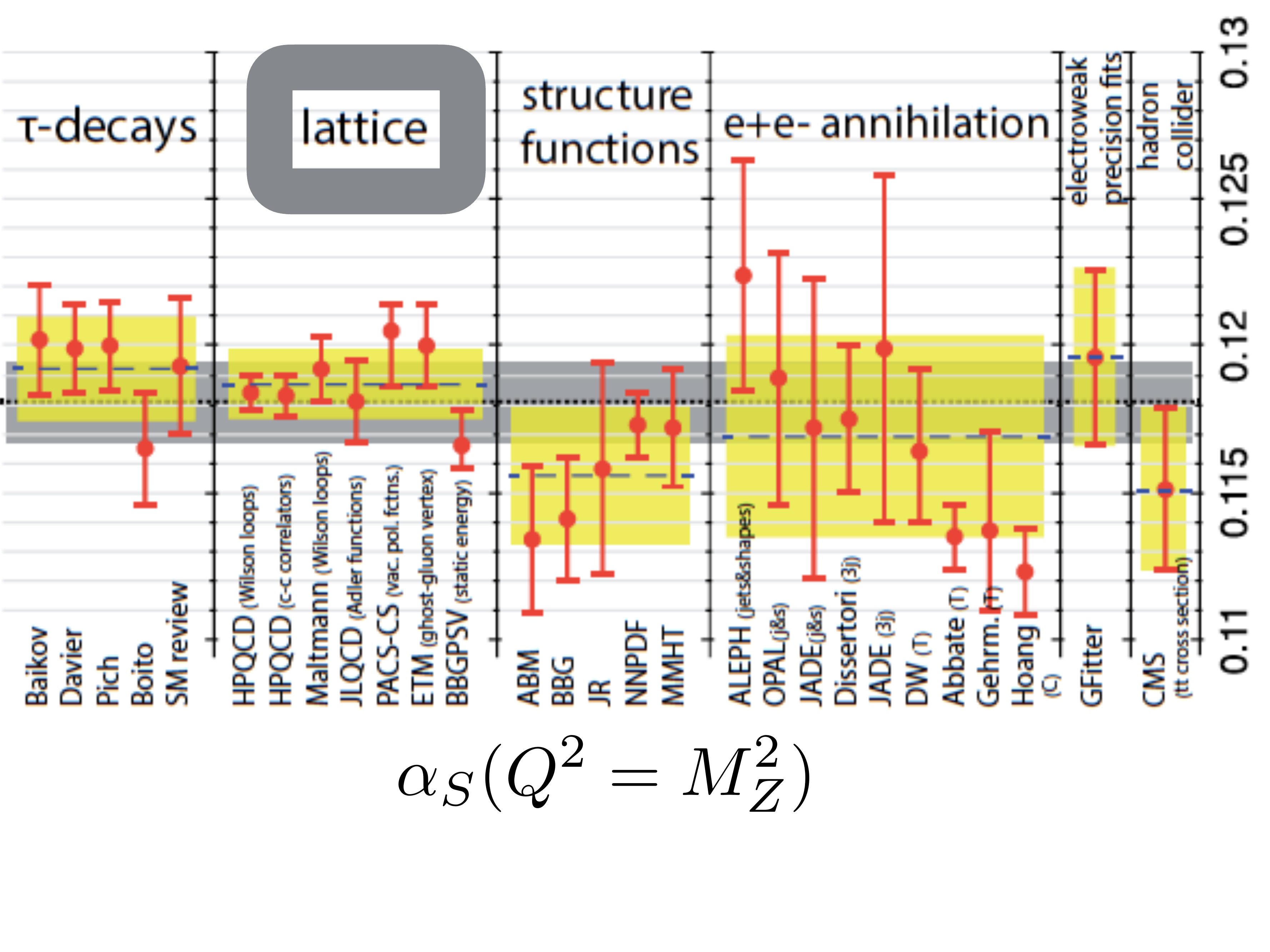}}
\vspace{-1cm}
\caption{Values\cite{alphas} of $\alpha_s(M_Z)$.}
\label{fig-alphasMZ}     
\end{figure}

The values of $\alpha_s$ at $Q^2=M_Z^2$, extracted in various ways, are summarized
in Fig.~\ref{fig-alphasMZ}, the references cited therein can be found in Ref.~\refcite{{alphas}}.
It is somewhat surprising and gratifying for theorists
that the precision of the lattice determinations competes so favorably with
that of the experimentally ``assisted'' ones.

\section{Back to the past: the irony of scaling in neutrino scattering}

The consensus that the observed scaling deviations smelled of QCD was
not triggered by theorists, but by an analysis of neutrino data by
Don Perkins {\it et al.}~\cite{Don}. This test of QCD was anything but
severe. One reason is the neglect of higher twists\cite{LM1,LM2}.
Furthermore it is not possible, event by event, to measure the 
neutrino energy. Thus, in an unintended implementation of Bloom--Gilman
duality, a measured structure function, $F_\nu(x,Q^2)$, is significantly blurred in $x$ and $Q^2$.
This erases, at the
low $Q^2$ of a good fraction of the data, the 
resonance bumps that must be there, as in Fig.~\ref{fig-BGD}, the total $\nu$
cross section increased linearly with $E_\nu$
 and was related to the naive, constituent-quark expectation from electroproduction by
a famous 18/5 factor, relating weak to electromagnetic quark charges.

Had the energy resolution of neutrino experiments
been as good as that of
their electron-scattering counterparts, the cross section rise would not have been linear at all,
the nucleon-resonance bumps in the structure functions would have been clearly
visible, and the data analysis would have had to be quite different.
With a pinch of poetic license one could assert that, early on, many
concluded that QCD was quite precise, but only  because the data were not.

\section{The QCD evolution of structure functions}

\begin{figure}[b]
\includegraphics[width=12cm]{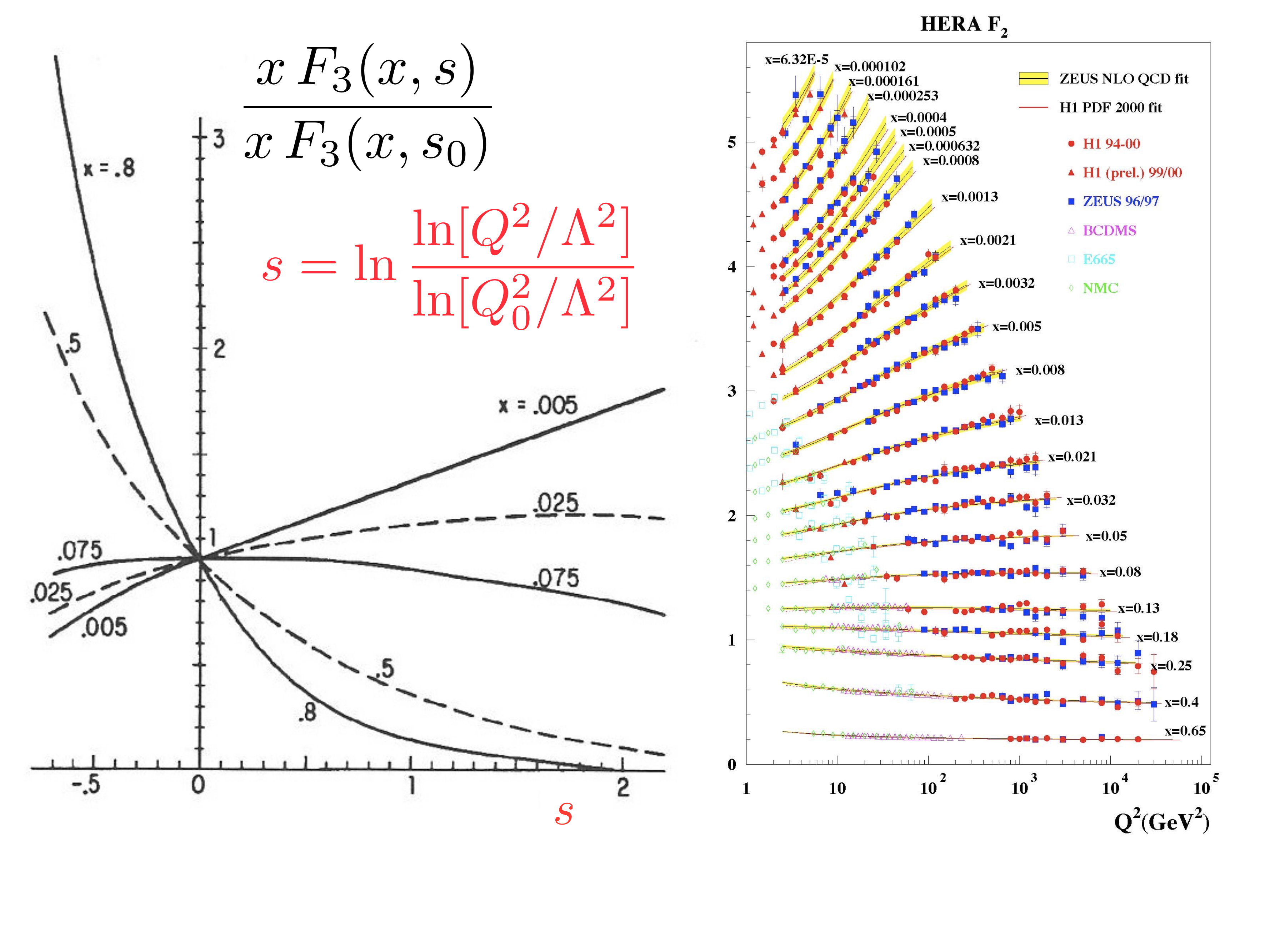}
\vspace{-1cm}
\caption{Left: Evolution of a normalized
$\nu$ structure function $F_3(x,s)/F_3(x,0)$ at fixed 
$x$. The trend has been corroborated in detail by a
multitude of experiments (and theorists). Right: HERA electron-scattering data.}
\label{fig-SFS}     
\end{figure}

Elaborating on work by Giorgio Parisi \cite{Giorgio},
Georgi, Politzer and I explicitly worked out the $Q^2$ evolution of structure functions at 
fixed $x$, large $Q^2$ (for which $\xi\simeq x)$ \cite{DGP1}. The simplest results 
are for $x\,F_3(x,Q^2)$ (the C-odd neutrino scattering structure function), shown in 
Fig.~\ref{fig-SFS}. A compilation of theory and HERA data for the electron scattering
$F_2(x,Q^2)$ is shown in  the same figure. 

The results shown in the left panel of Fig.~\ref{fig-SFS}
were to become heavily used...~and
systematically referenced to authors of later papers.
 While yowling, I plead guilty to having learned much later that
the simple underlying physics had been understood elsewhere: the
renormalization-group \cite{SP,GML}
picture of seeing partons within partons was drawn by
Kogut and Susskind  \cite{KS}, the ``physical gauge'' diagrammatic image of
a parton dissociating into others is due to Lev Lipatov
 \cite{Lip}, and its vintage QED analogue is none else than the
Weisz\"aker--Williams equivalent-photon approximation  \cite{WW}.

As it is extremely well known the QCD evolution became a hit 
with the 1977 publication of {\it Asymptotic freedom in parton language} by Guido Altarelli and 
Giorgio Parisi \cite{AlPa} (AP), which significantly simplified matters by avoiding the explicit
use the Mellin transforms of structure-function moments and directly using, instead,
quark and gluon parton density functions (PDFs). Discussing the relation of this paper
with the ones in the previous paragraph, Giorgio taught me something very wise.
To wit, one should compare Columbus to people having previously set foot in America,
such as the Vikings, not to speak of the Amerindians... to conclude that the important
thing concerning discoveries is not to be the first, but the last.

With time, ``the West" discovered that papers somewhat similar to AP's were published in the
Soviet Union. One, also in 1977, by Dokshitzer  \cite{YuLD},  and a much earlier one 
by Gribov and Lipatov \cite{VGLL}. The QCD evolution of structure functions and parton PDFs thus
became synonymous to DGLAP\footnote{The article by Gribov and Lipatov predates the
discovery of asymptotic freedom, does not deal
with composite hadrons and is not a
predecessor of AP in the same sense as the other articles I have cited. The article
of Dokshitzer is a close contemporary of AP (they were received by the 
respective journals one week apart)
but does not contain the simple AP equations.}.

\section{Progress in deep inelastic scattering}

Once again, in this section, I rely heavily on the contribution by Bethke, Disertori and Salam
in the Review of Particle Properties \cite{BDS}. In the customary notation, the
 parton model statement of Bjorken scaling is:
\begin{equation}
F_2(x,Q^2)=x \sum_q e_q^2\, f_{q/p}(x),\;\;\;
F_L(x,Q^2)=0,
\end{equation}
where the longitudinal structure function vanishes for point-like spin 1/2 constituents such as quarks.
The full QCD perturbative result is:
\begin{equation}
\mu_F^2\,{\partial f_{q/p}(x)\over \partial \mu_F^2}=
\sum_j{\alpha_s(\mu_F^2)\over 2\,\pi}
\int_x^1{dz\over z}P_{i\leftarrow j}(z)\,f_{q/p}\left({x\over z},\mu_F^2\right)
\label{eq:AP1}
\end{equation}
where $P_{i\leftarrow j}(z)\,f_{q/p}$ are the AP ``splitting functions" and
\begin{equation}
F_2(x,Q^2)=x\sum_{n=0}^\infty {\alpha_s^n(\mu_F^2)\over(2\,\pi)^n}\!
\sum_{i=q,g}\int_x^1 \! C_{n,i}^{(n)}(x,Q^2,\mu_R^2,\mu_F^2)\,
f_{i/p}\left({x\over z},\mu_F^2\right)+{\cal{O}}\!\left(\Lambda^2\over Q^2\right)\!,
\label{eq:AP2}
\end{equation}
 where the unspecified term reflects higher-twist corrections. The coefficient functions
 $C_{n,i}^{(n)}$ depend on two arbitrarily chosen scales, a 
 renormalization scale and a ``colinear" factorization scale: a boundary between 
 parton emissions taken care of by the coefficient functions and those included in
 the splitting functions.
 
 Since $F_2(x,Q^2)$ in Eq.~(\ref{eq:AP2}) is an observable, the dependence on $\mu_R^2$
 and $\mu_F^2$ should in principle disappear (in the absence of renormalon
 contributions) from a completely summed series, or at
 least become weaker and weaker as more terms are added. Not currently close to this ideal
 situation, the practice consists in defining a ``theoretical uncertainty" as the range of results
 in the brackets $\mu_R/2<\mu_F<2\, \mu_R$; $\mu_F/2<\mu_R<2\, \mu_F$. It is not clear
 how to do any better.

\begin{figure}[b]
\includegraphics[width=11cm]{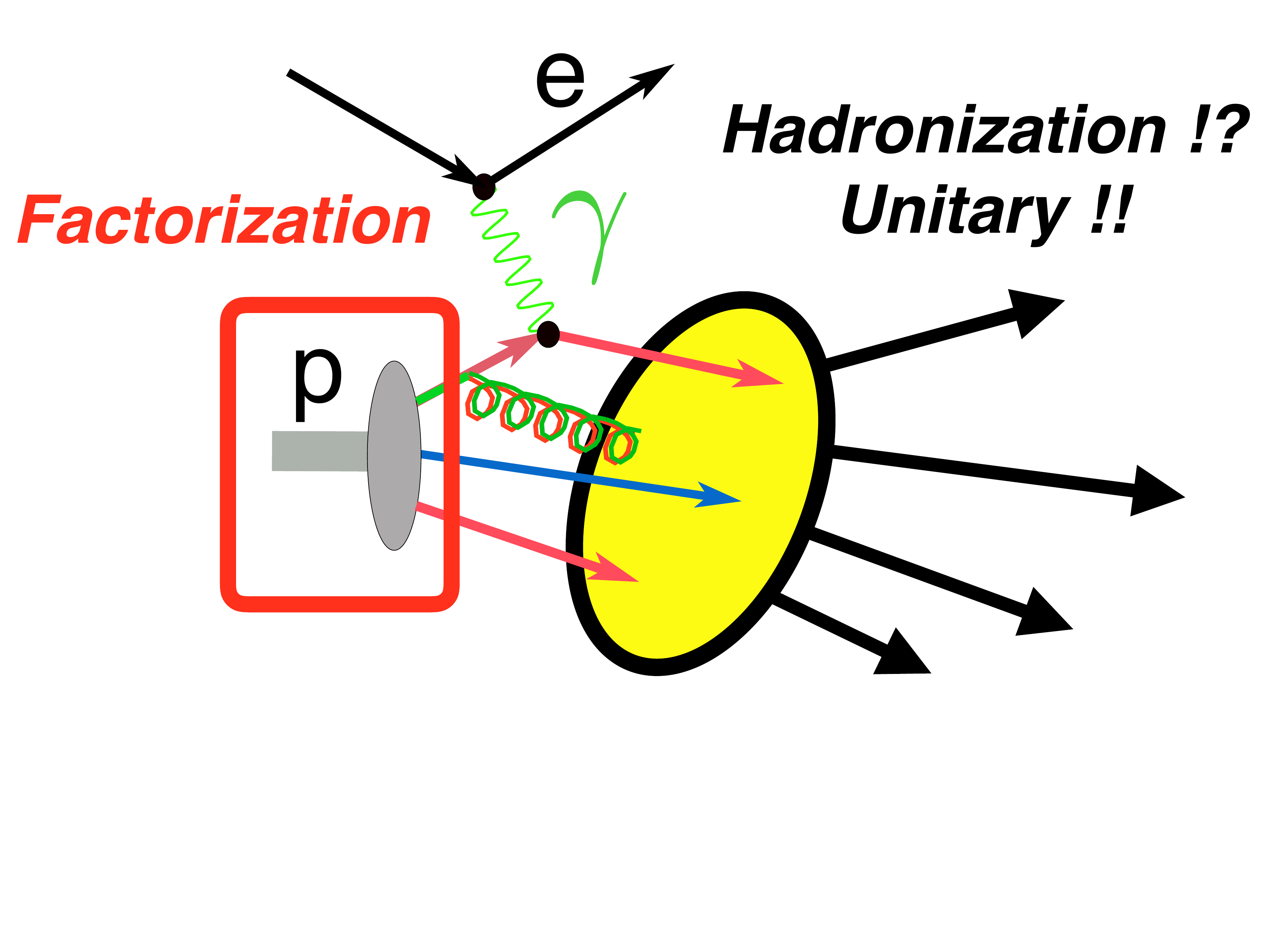}
\vspace{-2cm}
\caption{Factorization and hadronization in DIS.}
\label{fig-DeepInel}     
\end{figure}

An sketch of deep inelastic scattering (DIS) is shown in
Fig.~\ref{fig-DeepInel}, which illustrates the concept of the {\it factorization} of the process
into two steps. The first, involving the nucleon's constituents, occurs at a ``short distance" 
of ${\cal{O}}(1/\sqrt{-Q^2})$. The second is the {\it hadronization} of the scattered and {\it spectator}
partons, occurring at inter-parton distances of ${\cal{O}}(1/\Lambda)$. Factorization is the
hypothesis that the short- and long-distance processes do not interfere. This is augmented by the
{\it unitarity} argument that the semi-final-state partons have no choice but to end-up --with 100\%
probability-- confined within the outgoing hadrons.

Collins, Soper \& Sterman \cite{CSS} have analized in great detail the extent to which the
previous paragraph's arguments are defensible. They find them to be correct, even beyond
leading twist, in $(\Phi^3)_6$, the 6D renormalizable asymptotically free theory of scalars with
a cubic self-interaction. But for realistic processes in the Standard Model this is not the case.
Not surprisingly, the best understood process is DIS \cite{Guz,JSS,Ham},
even for higher-twist contributions from multiparton processes \cite{JSS,HK}.

Another item of interest concerns photons as partons. Progress has been made from an early
{\it The photon constituency of protons}\cite{YoV} to a decisive
{\it How bright is the proton?}\cite{Manohar}. 

\section{Drell-Yan-like processes}	

The production in $pp$ collisions of a single intermediate vector boson, or a couple of them,
are amongst the processes akin to the original ``Drell-Yan" process $pp\to e^+e^-+X$. In analogy
with Eq.~(\ref{eq:AP2}), the cross section for single-$W$ production, for instance, is written as:
\begin{eqnarray}
&&\;\;\;\;\;\;\;\;\;\;\;\;\;\;\;\;\;\;\;\sigma(h_1h_2\to W\,+\,X)= \nonumber\\
&&\sum_{n=0}^\infty \alpha_s^n(\mu_R^2)
\sum_{i,j}\int dx_1dx_2 f_{i/{h_1}}(x_1,\mu_F^2)
f_{j/{h_2}}(x_2,\mu_F^2)\nonumber\\
&&\;\;\;\;\;\;\times \; \hat\sigma_{ij\to W+X}^{(n)}
(x_1 x_2 s,\mu_R^2,\mu_F^2)+{\cal{O}}
\left({\Lambda^2\over M_W^4}\right).
\end{eqnarray}
Here, the caveats concerning the use of the two scales $\mu_F$ and $\mu_R$ are the same as 
the ones we have already commented upon.
What is worse, for these processes
the infrared divergencies in higher-twist contributions do not cancel beyond
the one-loop level\cite{GrLi,Lip,AlPa,YuDo,CFP}. 

\begin{figure}[b]
\includegraphics[width=11cm]{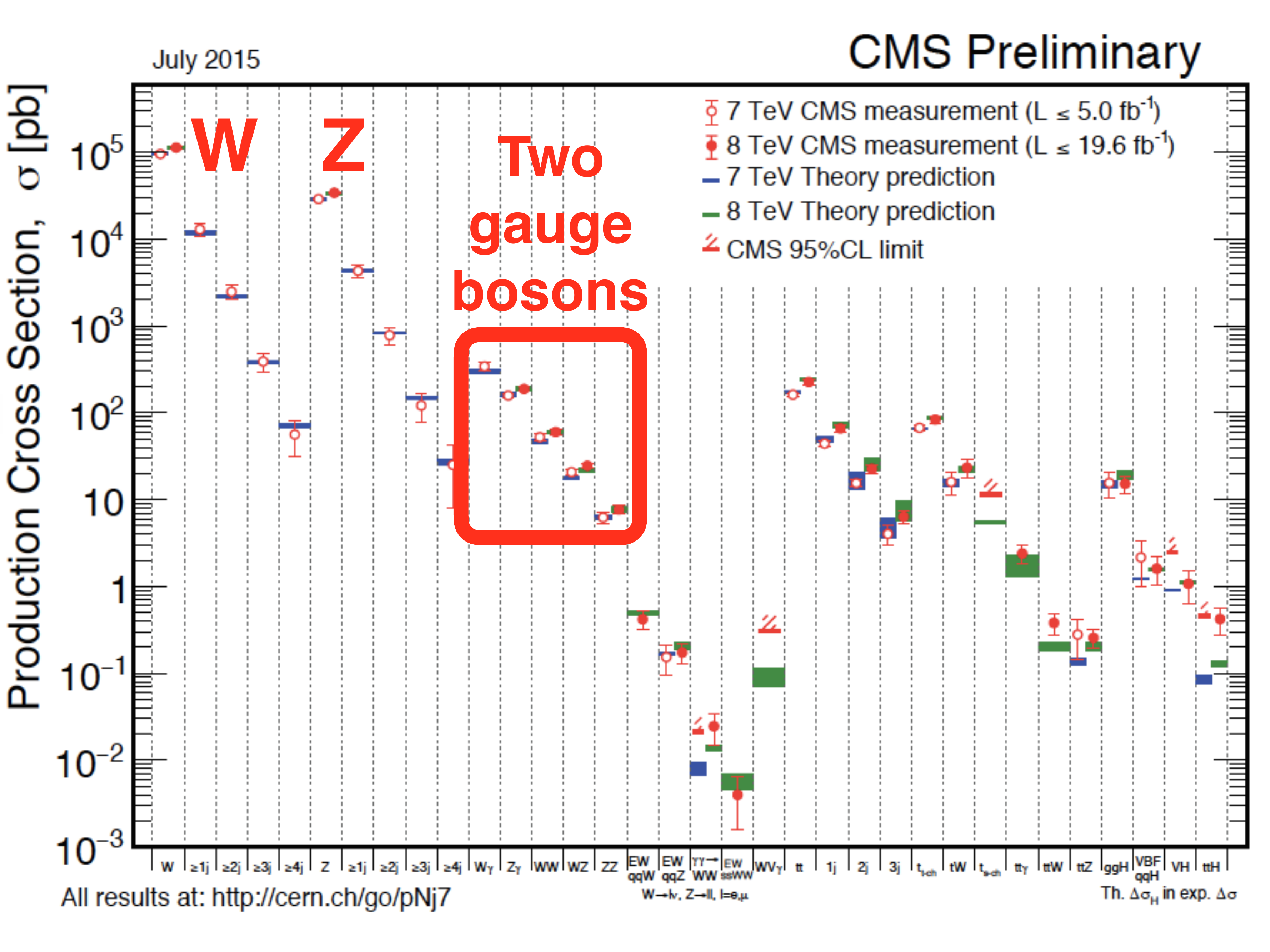}
\caption{Comparison of CMS data with theory. Notice the results for single and even double gauge boson 
production.}
\label{fig-DYCMS}     
\end{figure}

\begin{figure}[b]
\includegraphics[width=11cm]{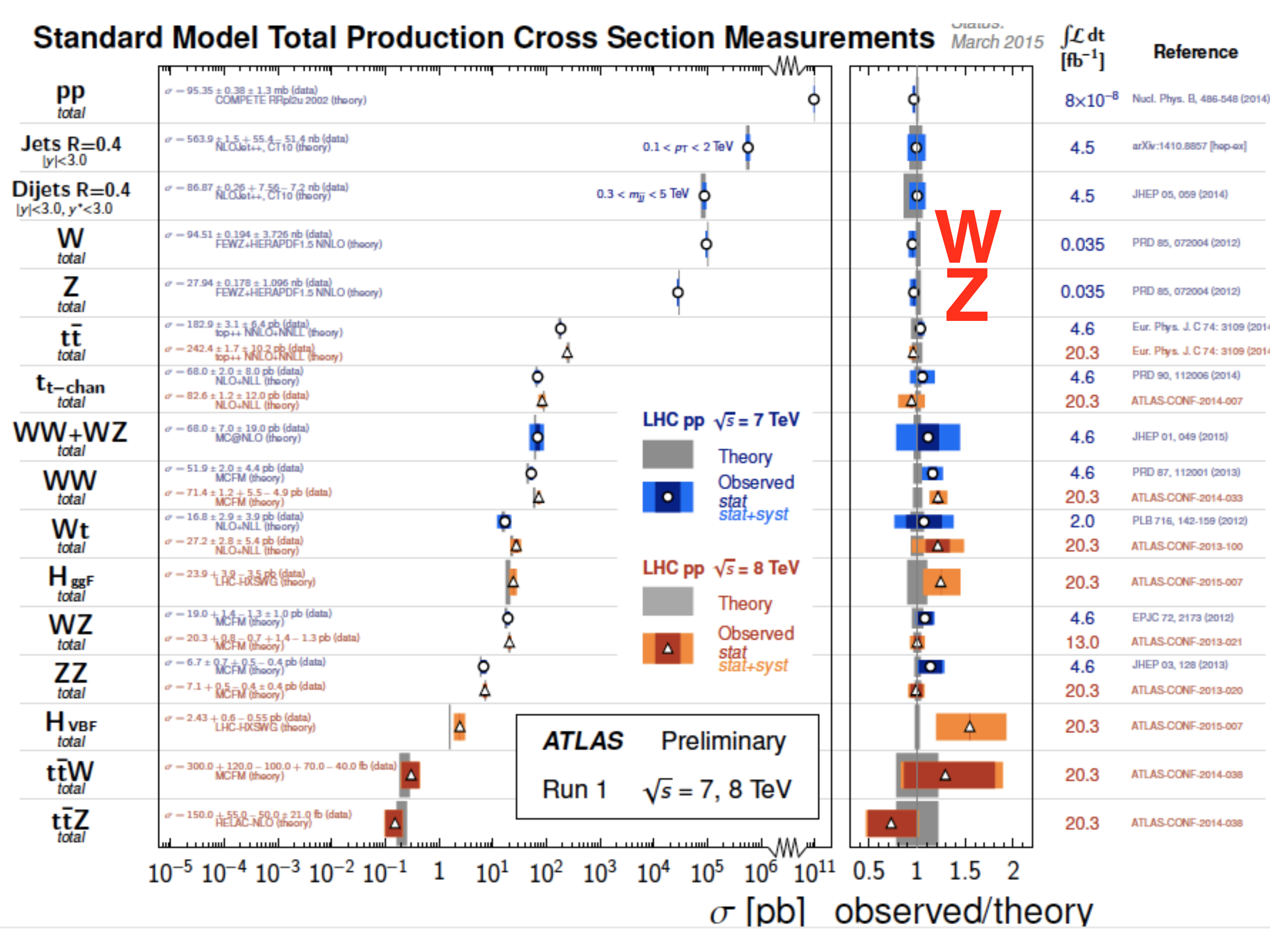}
\caption{Comparison of Atlas data with theory. Notice the results for single and even double gauge boson 
production.}
\label{fig-DYAtlas}     
\end{figure}

Quite welcome --but somewhat surprising-- is the fact that, in spite of the above admonitions, theory
and data agree to their current level of precision. This is shown for CMS data in Fig.~\ref{fig-DYCMS}
and for Atlas data in Fig.~\ref{fig-DYAtlas}. The perturbative NNLO calculations, as well as the data
are reaching a precision at which it will no longer be safe to ignore Next to Leading Twist effects.

\section{The November Revolution}

\begin{figure}[b]
\centerline{\includegraphics[width=10cm]{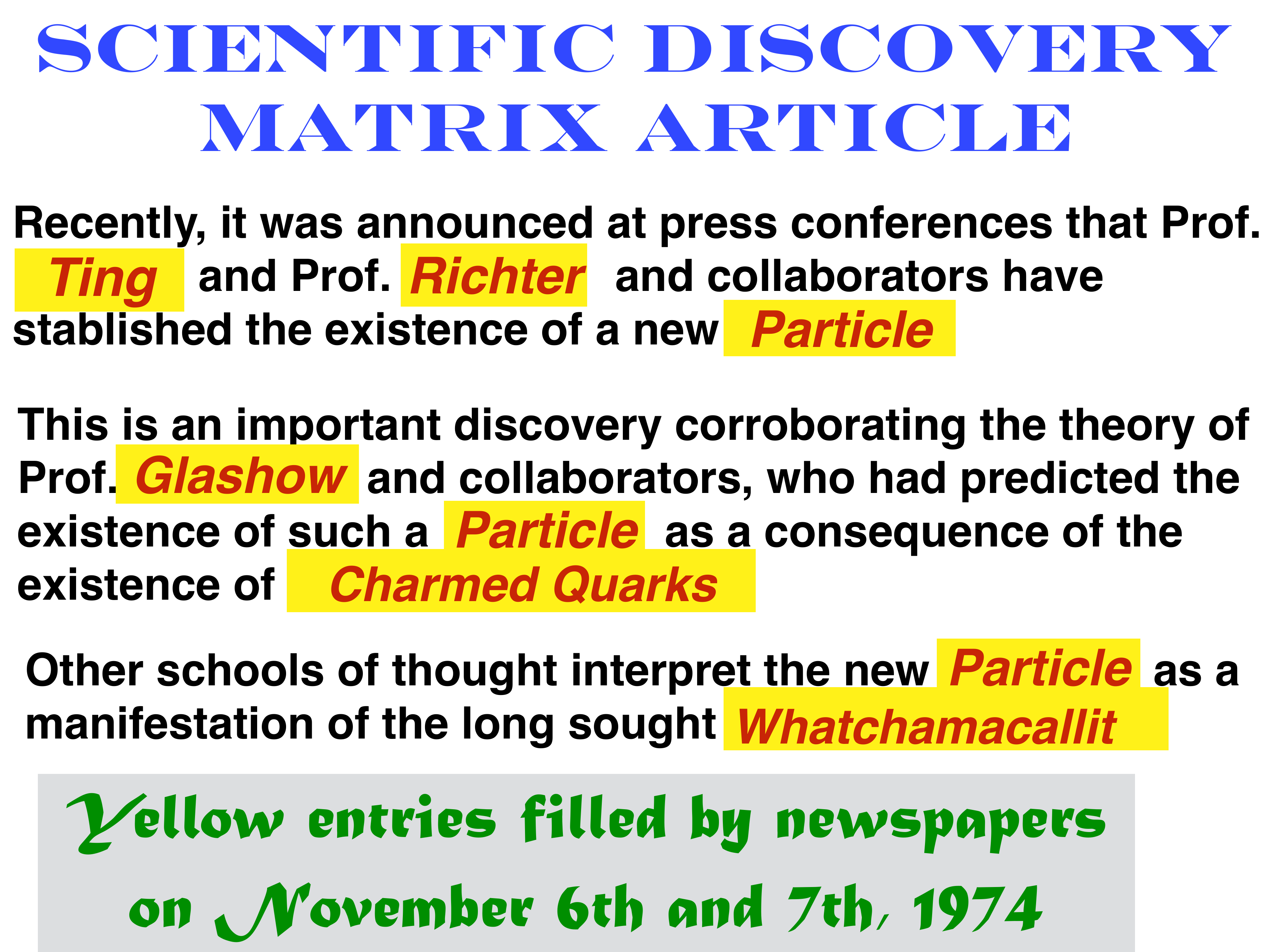}}
\caption{The Scientific Discovery Matrix Article. The November 74 Revolution; fully filled version.}
\label{fig-SciDis}     
\end{figure}

In the 1970's the pace of discovery was so fast that (lazy) journalists decided to prepare
a ``matrix article'', wherein only the details of each specific discovery had to be filled in
at the last minute. The NYT matrix article and its filling in November 1974 are shown
in Fig.~\ref{fig-SciDis}.

\begin{figure}[b]
\centerline{\includegraphics[width=11cm]{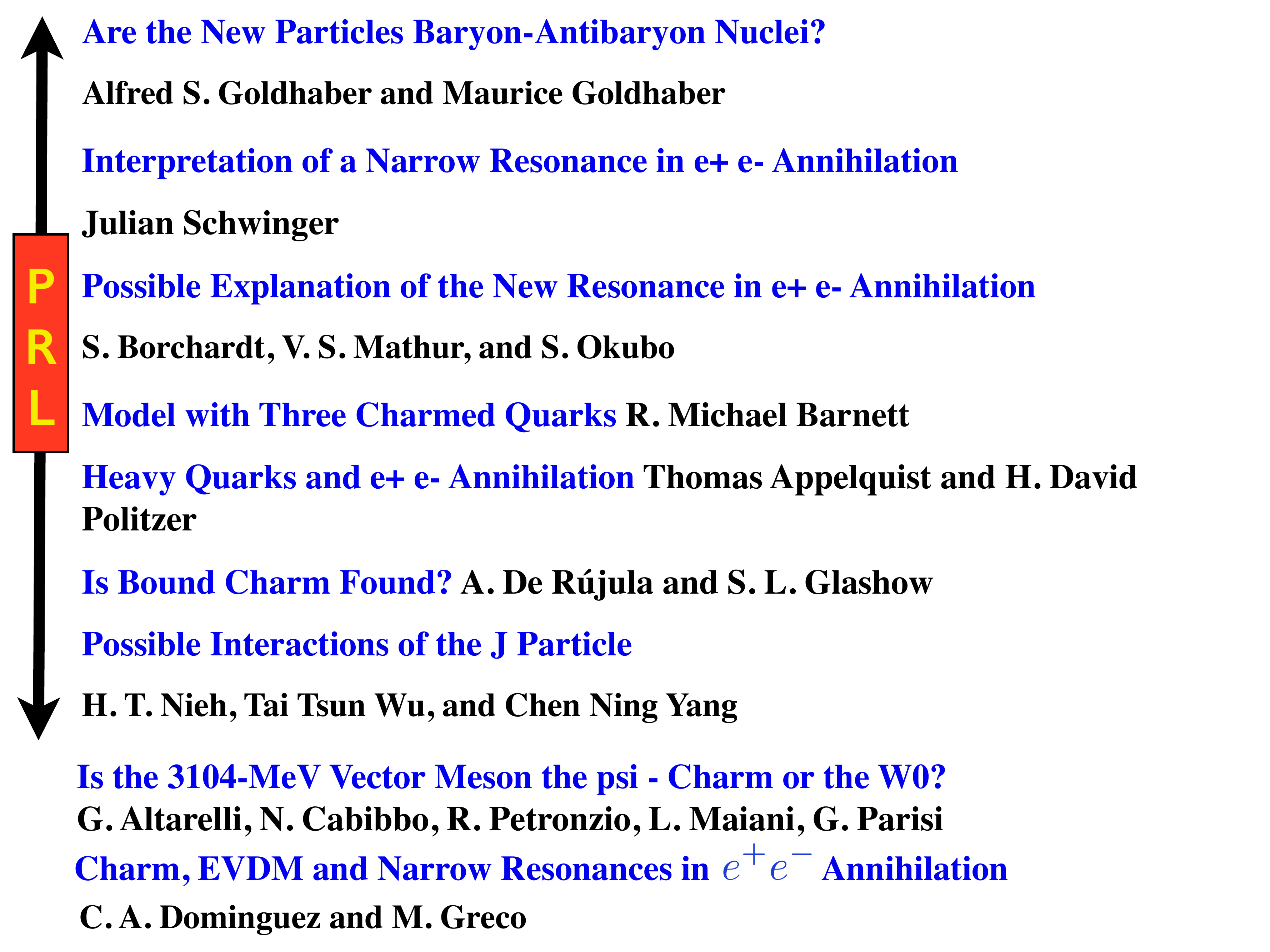}}
\caption{Immediate interpretations of the $J/\psi$, with their titles.
PRL is Phys. Rev. Lett. {\bf 34}, Jan.~6th, 1975.
The last two papers\cite{Romans,Mario} are in Lett. Nuovo Cim.}
\label{fig-Interpretations}     
\end{figure}

Ten days of November 1974 shook the world of physics. 
Something wonderful and {\it almost}  \cite{AP} unexpected was to see the
light of day:  a  very discreetly charmed particle
 \cite{J,Psi}, a hadron so novel that it hardly looked like one. Two score and five
years later it is not easy to  recall  the 
collective ``high'' in which this discovery\footnote{The ``usual Russian suspects", so often
ahead of Westerners, did not on this occasion have the time to contribute. This is in
spite of the fact that one of them obtained permission from the Soviet authorities to
give me a phone call to catch up on the news.}, and others to be made
in the two consecutive years, submerged the particle-physics community.
In my opinion, a detailed account that reflects well
the mood of the period is that by Riordan  \cite{Riordan}. In a
nutshell, the standard model arose from the ashes of the November
Revolution, while its competitors died honorably on the battleground.
A couple of survivors and many of the casualties are shown in Fig.~\ref{fig-Interpretations}.
All of them but the last two were published, {\it unrefereed,} in the same issue of PRL.

Burt Richter was also caught in the November avalanche. 
On a short visit to Harvard, and with a healthy disrespect for theory, Burt told
us that the electron spent some of its time as a hadron. 
In answer to a question by Applequist, he explained that sufficiently narrow
resonances would escape detection in $e^+e^-$ colliders. Nobody around 
was aware of the possibility of catching the
devil by its radiative tail (the emission of photons by the colliding
particles widens the observed resonance on its $\sqrt{s}>M$ side). 
Our vain discussions came to an abrupt end; 
a rather urgent call summoning Burt back to SLAC delivered us from his 
scorn for theorists. 

The experimental papers of the Brookhaven group
led by Samuel Ting and the SLAC one led by Richter were sent for
publication within an interval the reader should by now be familiar with: one day.
Their statistical evidence for a discovery was more than sufficient to remind
one of a contemporary dictum by Val Telegdi: {\it If you need statistics to prove your
point, you may not have one.}

In the early Fall of '74, Tom Appelquist and David Politzer 
had been looking leisurely at how asymptotic freedom could imply a
positronium-like structure for the $c\bar c$ bound states of a
charmed quark and its  anti-particle. In those days, both QCD and
charm were already fully ``established'' at Harvard.
Since Americans are often
short of vocabulary, my first contribution to the
subject was to baptize their toy {\it charmonium}. David and Tom's
first charmonium spectrum was so full of  Coulomb-like peaks  that they
could not believe it themselves. They debated the problem long enough
for the experimental avalanche to catch up with them. It was  a heavy 
price to pay for probity.

For an object of its mass, the $J/\psi$ is four orders of magnitude
narrower than a conventional hadron resonance, and one order of magnitude
wider than a then hypothetical weak intermediary. It could not be either.
 Of the multitude of theoretical papers of Fig.~\ref{fig-Interpretations},
 only two attributed the narrow width to asymptotic
freedom, one by Tom and David  \cite{AP}, who had intuited the whole
thing before, the other one by Sheldon Glashow and me  \cite{DeRG1}. I
recall Shelly storming the Lyman/Jefferson lab corridors with the notion of the
feeble three-gluon hadronic decay of the $J^P=1^-$ {\it
orthocharmonium} state, and I remember Tom and David muttering:
``Yeah''. Our paper still made it to the publishers in the auspicious
November, but only on the 27$^{\rm th}$, a whole week after the
article of our Harvard friends. The paper by Cesareo Dominguez and Mario Greco\cite{Mario}
also singled out charmonium as the interpretation.

Glashow and I did a lot in our extra week  \cite{DeRG1}. {\it Abusus non tollit
usum} (of asymptotic freedom) we related the hadronic width of the
$J/\psi$ to that of $\varphi \to 3\,\pi$, to explain why this hadronic resonance
 {\bf had to be} so narrow. For its dominant decay into hadrons, we estimated a width:
 \begin{equation}
 \Gamma={3\over 2}\,{M_{J/\Psi}\over M_\varphi}
 \left[{\alpha_s(M_{J/\Psi})\over \alpha_s(M_\varphi)}\right]^6\,\Gamma(\varphi\to 3\pi)
 =42\,{\rm keV},
 \label{width}
 \end{equation}
 with $\alpha_s(M_{J/\Psi})$ QCD evolved from the three-gluon estimate:
 \begin{equation}
 \alpha_s(M_\varphi)={3\over 2} \left[{
9\;\pi\;\Gamma(\varphi\to 3\,\pi) \over
10\,(\pi^2-9)M(\varphi)  } \right]^{1/6}.
 \label{width2}
 \end{equation} 
 It is fun to quote QCD results from the times they were so simple.
 
  \begin{figure}[b]
\centerline{\vspace{-.8cm}\includegraphics[width=11cm]{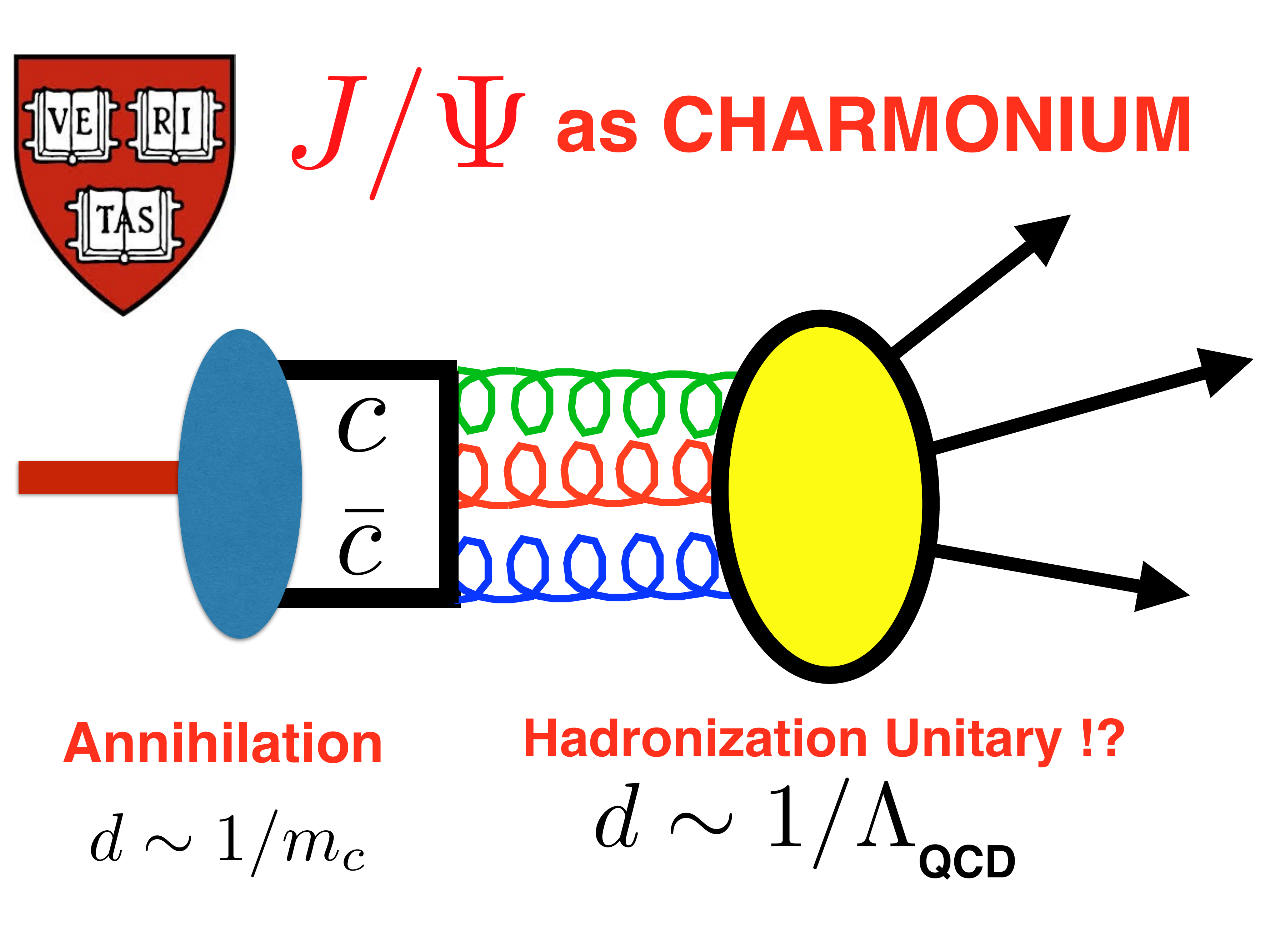}}
\caption{Annihilation of a $c\bar c$ pair and subsequent hadronization.}
\label{fig-JPsiwidth}     
\end{figure}

 The result of Eq.~(\ref{width}) is to be compared with the currently
 measured hadronic width, $\sim 59$ keV.  A rather good prediction, particularly considering
 that it depends on $\alpha_s(M_{J/\Psi})/ \alpha_s(M_\varphi)$ to an error-enhancing
 sixth power. Its well known basic input is shown
 in Fig.~\ref{fig-JPsiwidth}, involving the same considerations as the
 ones about Fig.~\ref{fig-DeepInel}.
 
 In Ref.~\refcite{DeRG1} we did also correctly estimate the
yields of production of truly charmed particles in $e^+e^-$
annihilation, $\nu$-induced reactions, hadron collisions and
photoproduction. Our mass for the $D^*$ turned out to be $2.5 \% $
off, sorry about that. We even discussed
mass splittings within multiplets of the
same quark constituency as {\it hyperfine}, a fertile notion. In
discussing paracharmonium ($J^P=0^-$) we asserted that {\it ``the
search for monochromatic $\gamma$'s should prove rewarding''}.
Finally, we predicted the existence of $\psi '$, but this time it was
our turn to be overtaken by the pace of discovery. 

\section{Charmonium spectroscopy}

I have a few vivid printable recollections of the times I am
discussing. One  concerns the late night in which the existence of
$P$-wave charmonia hit my head: we had been talking about $L=0$
states without realizing (we idiots!) that a bunch of $L=1$ charmonia
should lie between $J$ and $\psi '$ in mass. Too late to call Shelly,
 I spent hours guessing  masses and estimating the obviously
all-important $\gamma$-ray transition rates. At a gentlemanly morning
hour I rushed on my bicycle to Shelly's office, literally all the way
in, and attempted to snow him with my findings. I was speechless, out of breath
and wits. Shelly profited
to say: ``I know exactly what you are trying to tell me, there are
all these $P$-wave states etc., etc.'' He had figured it all out at
breakfast. I hated the guy's guts.

\begin{figure}[b]
\centerline{\includegraphics[width=12cm]{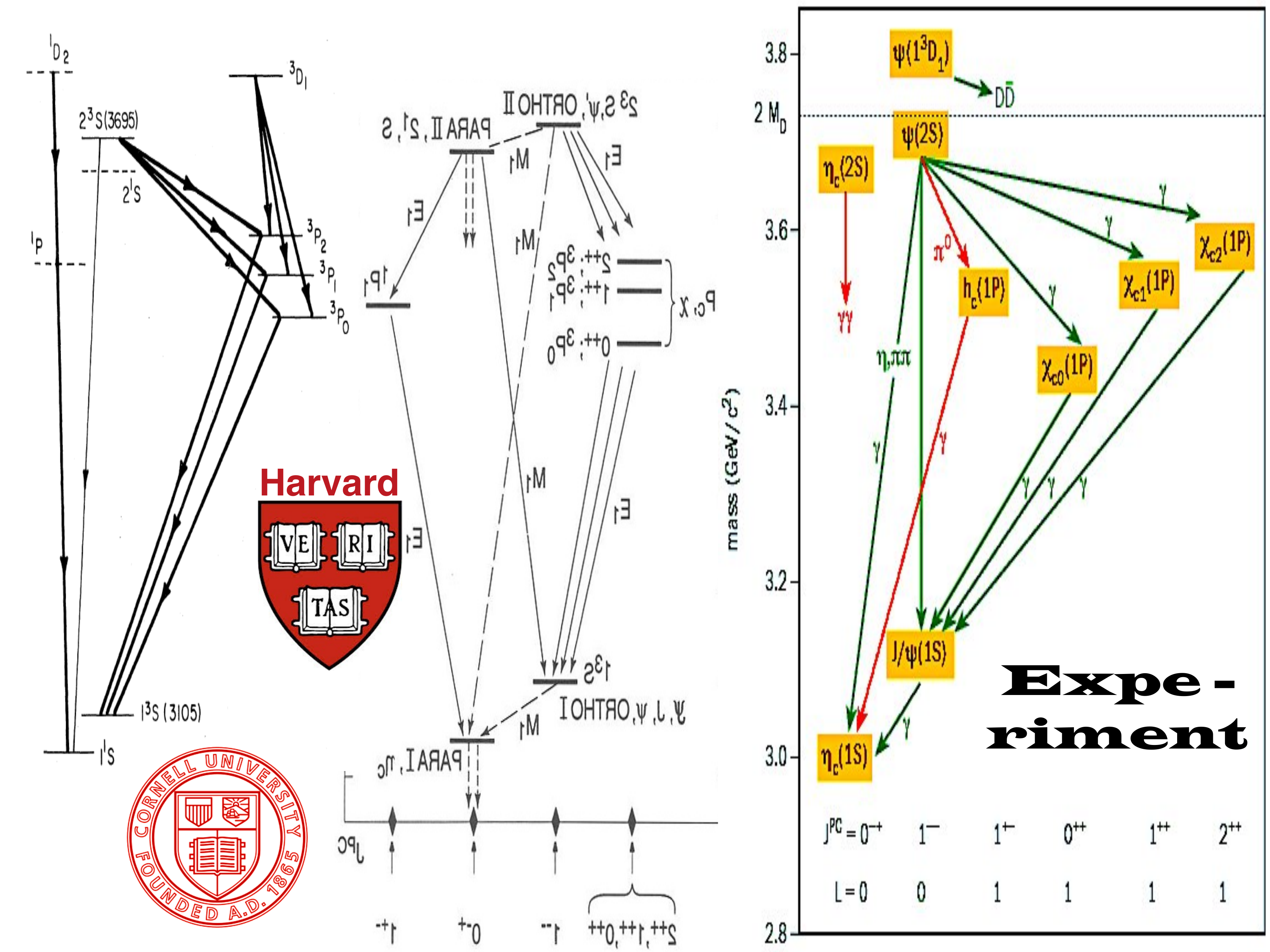}}
\caption{The spectra of charmonia. From left to right: Cornell's (squeezed), Harvard's (mirror reflected) 
and  observed (with the inclusion of some non-radiative decays).}
\label{fig-Charmonia}     
\end{figure}

In no time, David and Tom gathered forces with Shelly and me to produce
an article  \cite{ADGP} on {\it Charmonium spectroscopy.} Physical
Review Letters  was fighting its usual losing battle against
progress (in nomenclature, $\smile$) and did not accept the title. 
Neither did PRL accept a similar title by our friendly Cornell competitors \cite{Cornell}.
The predicted spectra and the current experimental situation are shown
in Fig.~\ref{fig-Charmonia}.

We, {\it the Crimson}, estimated the energy levels as ``half-way" between those
of a Coulomb and a harmonic oscillator potential. Indeed, a linear potential
--adequate for confined $c\bar c$ states-- is in its dependence on distance
somewhat half-way. Our Cornell
friends, {\it the Carnelian}, borrowed a linear-potential program from Ken Wilson
and got similar results. Except for the all-important $\gamma$-ray transition
rates, for which the Carnelian predictions were much better than ours.

\section{Hadron masses in a gauge theory}

Early in 1975, Georgi, Glashow, and I wrote a paper \cite{DGG} whose style
reflects how high we rode, as well as how unorthodox QCD still was.
But for the added parentheses, here is how it began:

{\it Once upon a time, there was a controversy in particle physics.
There were some physicists who denied the existence of structures
more elementary than hadrons, and searched for a self-consistent
interpretation wherein all hadron states, stable or resonant, were
equally elementary} (the bootstrap). {\it Others, appalled by the teeming democracy of
hadrons, insisted on the existence of a small number of fundamental
constituents} (quarks) {\it and a simple underlying force law} (QCD). {\it In terms of these
more fundamental things, hadron spectroscopy should be qualitatively
described and essentially understood just as are atomic and nuclear physics.}

To the non-relativistic  quark model, we
added chromodynamic interactions
entirely analogous to their electrodynamic counterparts. We shall see that to this day
it is not totally clear why the ensuing predictions were so good. Our paradigmatic
result was the explanation of the origin and magnitude of the
$\Sigma^0$--$\Lambda$ mass difference. The two particles have the same
spin and quark constituency, their mass difference is a {\it
hyperfine} splitting induced by spin--spin interactions between
the constituent quarks. A little later, the ``MIT bag'' community published their 
relativistic version  \cite{MIT} of the same work.

In  Ref.~\refcite{DGG} we also predicted the masses of all ground-state
charmed mesons and baryons and (me too, I'm getting bored with this)
we got them right on the mark. Predictions based on an
incredible SU(4) version of the Gell-Mann--Okubo SU(3) mass formula,
and also the more sensible bag results, turned out to be wrong.
Only one person --indeed, again a Russian-- trusted a ``QCD-improved'' 
constituent quark model early enough to make predictions somewhat
akin to ours: none less than Andrei Sakharov  \cite{Sach}.

\subsection{Good News at Last}

While theorists faithfully ground out the phenomenology of QCD,
experimentalists persistently failed to find decisive signatures of our
Trojan horse: the charmed quark. At one point,  the upper limits on the
$\gamma$-ray transitions of charmonia were well below the theoretical
expectations. Half of the $e^+e^-$ cross-section above $\sqrt{s}=4$ GeV was
due to charm production, said we. Who would believe that experimentalists
couldn't tell?

In the winter of '75 we saw a lone ray of light. As Nick Samios
recalls in detail in Ref.~\refcite{NSamios}, 
a Brookhaven bubble-chamber group  \cite{Nick} pictured
a $\Delta S=-\Delta Q$ event, forbidden in a charmless world, and
compatible with the chain:
\begin{equation}
\nu_\mu\;p  \rightarrow \Sigma^{++}_c \mu^-\, ; \;\;\;
\Sigma^{++}_c  \rightarrow \Lambda^+_c\; \pi^+\, ;  \;\;\;
\Lambda^+_c  \rightarrow \Lambda\;\pi^+\;\pi^+\;\pi^- \; .\nonumber
\end{equation}
Two charmed baryons discovered
in one shot!  This was a
source of delight not only for us, but also for the experimentalists
involved. They deserve a reproduction of their event: Fig.~\ref{fig-NS}, and a quotation:

\begin{figure}[b]
\centerline{\includegraphics[width=11cm]{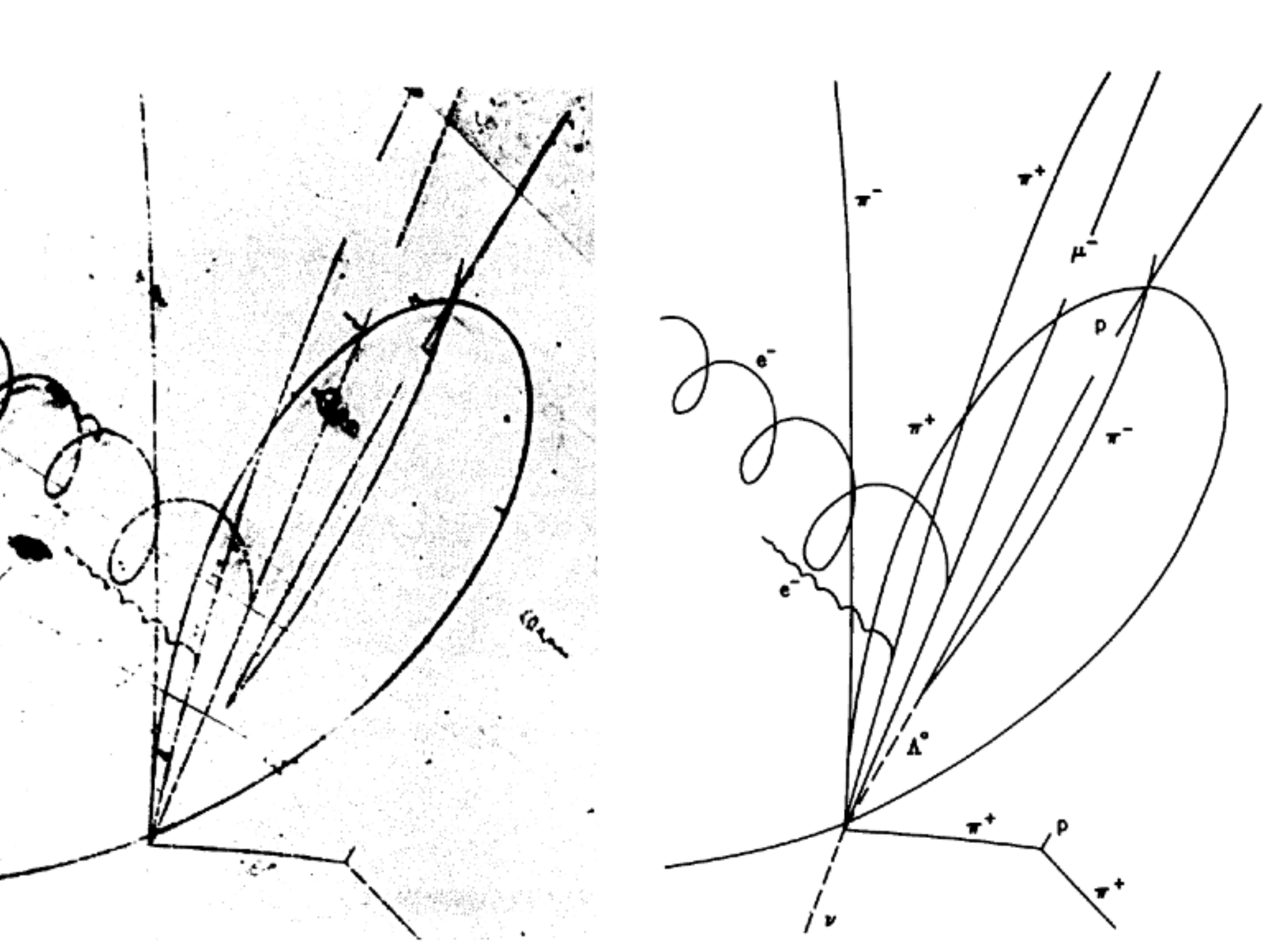}}
\caption{The Brookhaven doubly charming event \cite{Nick}.}
\label{fig-NS}     
\end{figure}

{\it The total recoiling hadron mass ($\Lambda\,\pi^+\pi^+\pi^-$)} [is] $2426\pm 12$ {\it MeV.}

{\it This mass is in reasonable} {\it agreement with the value predicted by De R\'ujula, Georgi
and Glashow for the lowest-lying charmed-baryon states of charge +2, $2420$ MeV}
($J^P={3 \over 2}^+,\; I=1,\;\Sigma_c^*$) ... {\it There are three $\pi^+$'s and thus three mass 
differences derivable form this event; these
are observed to be $166\pm15$ MeV, $338\pm12$ MeV, and $327\pm 12$ MeV. The first of
these differences is in remarkable agreement with the $160$ MeV predicted for the decay of
a spin-$1\over 2$ charmed baryon $\Sigma_c$ decaying into a charmed $\Lambda_c$.}

This is {\it almost} precisely the way I feel experimentalists should write papers. 
Only ``almost'' because
the agreement between $2426\pm 12$ and 2420 MeV seems to me to be a bit better
 than ``reasonable''.
 
What made QCD part of the now generally accepted Standard Model?
Even the most formal theorist or the most cable-connecting experimentalist
understands positronium and hydrogen. These objects are not so very different
from their QCD analogs, charmonium
and charmed particles. This may be why it took asymptotic freedom and a
fourth quark to have the Standard Model become the standard lore.

\section{Back to the present: charmed baryon masses}  

\begin{figure}[b]
\centerline{\includegraphics[width=12cm]{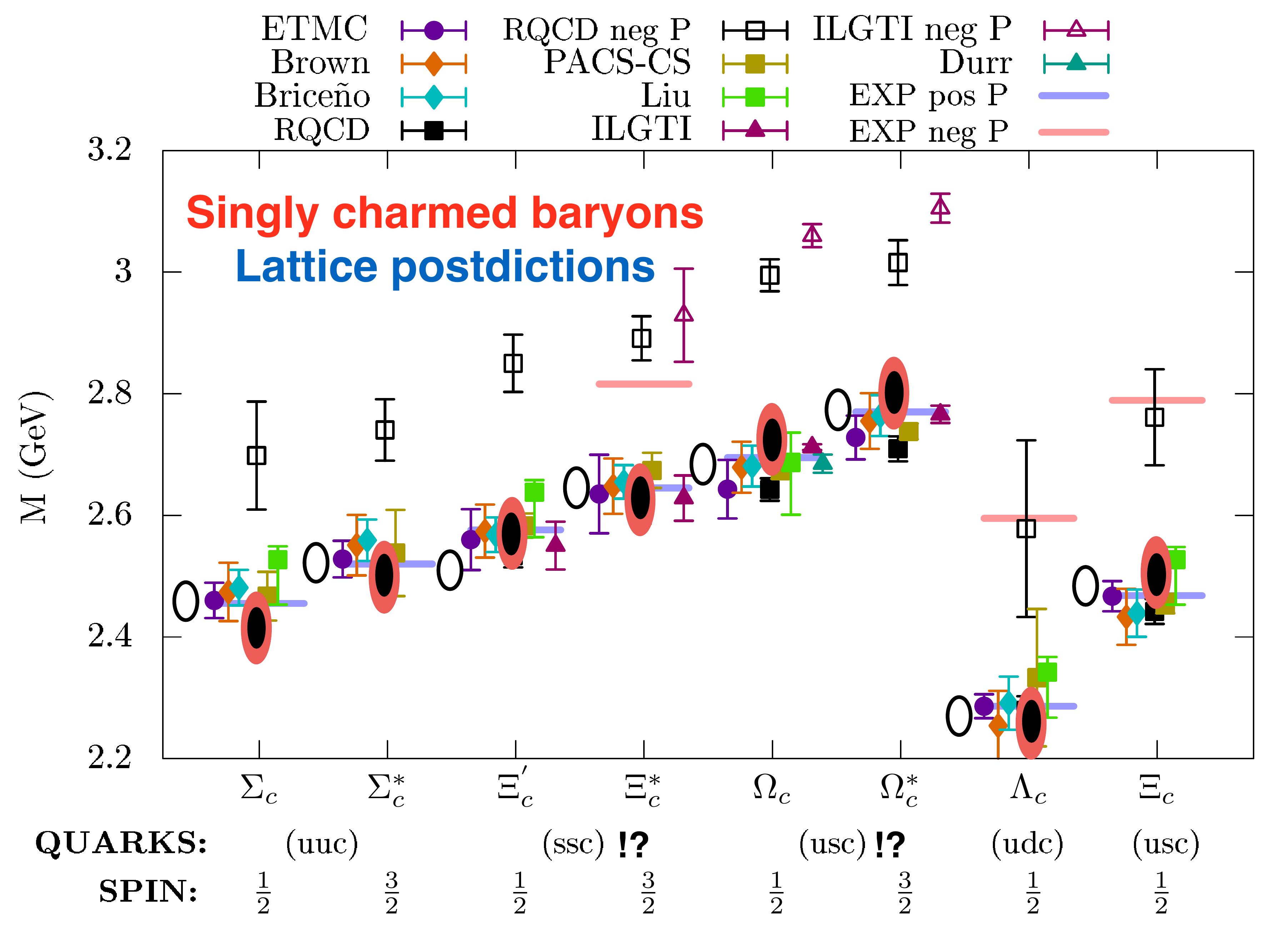}}
\caption{Charmed baryons. Blue traits: mass measurements for the positive parity
ones.  Red ellipses: the predictions\cite{DGG},
with their common uncertainty of $\pm 50$ MeV. Colored circles, squares, diamonds and triangles:
a compilation of unquenched lattice 
post-dictions \cite{PCB}. The labels (ssc) and (usc) are interchanged, as
in the original figure \cite{PCB}. Unfilled ellipses: results of an quenched QCD calculation \cite{Unq}
(their $\Xi_c$ result  is probably a typo.)} 
\label{fig-CBs}     
\end{figure}

By now the masses of many mesons and baryons have been precisely post-dicted
in lattice QCD. It is perhaps instructive to look at an instance:  charmed baryons.
This is done in Fig.~\ref{fig-CBs}, where a collection of lattice results \cite{PCB}    
is shown, along with the observed values (blue lines) for the positive parity baryons
we shall discuss. Also shown in the figure are the predictions of Ref.~\refcite{DGG},
made after the discovery of the $J/\psi$, but before that of open charm.
This limited information made us (over)estimate a common uncertainty of
$\pm 50$ MeV --reflected as the red ellipses-- around the central values of the masses.

The QCD-improved naive quark model predictions \cite{DGG} still compete
with the lattice results. The moral is that there is an element of ``truth'' in the
naive model: unresolved by a short-distance probe, a baryon consists of
three quarks (and some glue). This is strongly corroborated by the quenched lattice results\cite{Unq}
shown as unfilled ellipses in Fig.~\ref{fig-CBs}: adding $q\bar q$ does not
significantly modify the lattice predictions.

Presumably, if latticists could device a gauge-invariant way to characterize
a confined constituent-quark propagator, it would be found to peak at a
``constituent mass'' some $300$ MeV larger than a ``Lagrangian'' or chiral-model 
mass\footnote{This happens in attempts to describe confinement by methods based on
Bethe-Salpeter \cite{CG} or Schwinger-Dyson \cite{ACFP} equations. I am indebted to Pilar 
Hernandez and Carlos Pena for discussions on this topic.}.
Similarly, they might discover one gluon exchange dominance for the
mass differences between baryons of the same quark constituency but
different spin or quark-spin alignements.

 \section{Lattice predictions}

Two areas of QCD have 
witnessed an enormous progress over the years, as summarized, for instance, in Ref.~\refcite{SLS}.
One of them is the non-perturbative
first-principle lattice calculation of many relevant observables. 
These include exotic hadron masses and decays, glueballs,
form factors in $K,\,D,\,B,\,\Lambda_b$ and 
$\tau$ decays, moments of structure functions, $K\to \pi\pi$
amplitudes, the hadronic vacuum polarization contribution to $g_\mu-2$, long-distance contributions
to $K \leftrightarrow \bar K$ mixing and rare $K$ decays.

Not having ever been active on the subject, I shall give only one recent example
 of a lattice prediction that turned out to be very relevant. That is a result for
 $\vert V_{ub}/V_{cb}|$,  the ratio of two entries
in the CKM matrix and the corresponding unitarity triangle. 
To understand the data on the ratio of the decay rates $\Lambda_b\to p\,\mu^-\bar\nu_\mu$
and $\Lambda_b\to \Lambda_c\,\mu^-\bar\nu_\mu$, integrated over given ranges in 
$q^2$ (with $q$ the momentum transfer between the baryons) it appeared to be necessary
to invoke right-handed weak currents\cite{LHCb}, absent in the Standard Model. When
the matrix elements $\textless p | j^\mu | \Lambda_b\textgreater$ and 
$\textless \Lambda_c | j^\mu | \Lambda_b\textgreater$ for vector and axial currents, $j^\mu$
--required to extract $\vert V_{ub}/V_{cb}|$-- were calculated in the lattice\cite{DLM} the
problem disappeared. Yet another {\it Beyond the Standard Model} (BSM) result that goes down the drain.

\subsection{The $\Delta I=1/2$ rule, quite briefly}

The fly in the lattice ointment is the $\Delta I=1/2$ rule, an issue that lattice calculations
have not solved. But this is a different type of BSM result: 
it is a {\it Before the Standard Model} problem.

\section{Beyond $\mathbf{q\bar q}$ and $\mathbf{qqq}$}

\begin{figure}[b]
\centerline{\includegraphics[width=12cm]{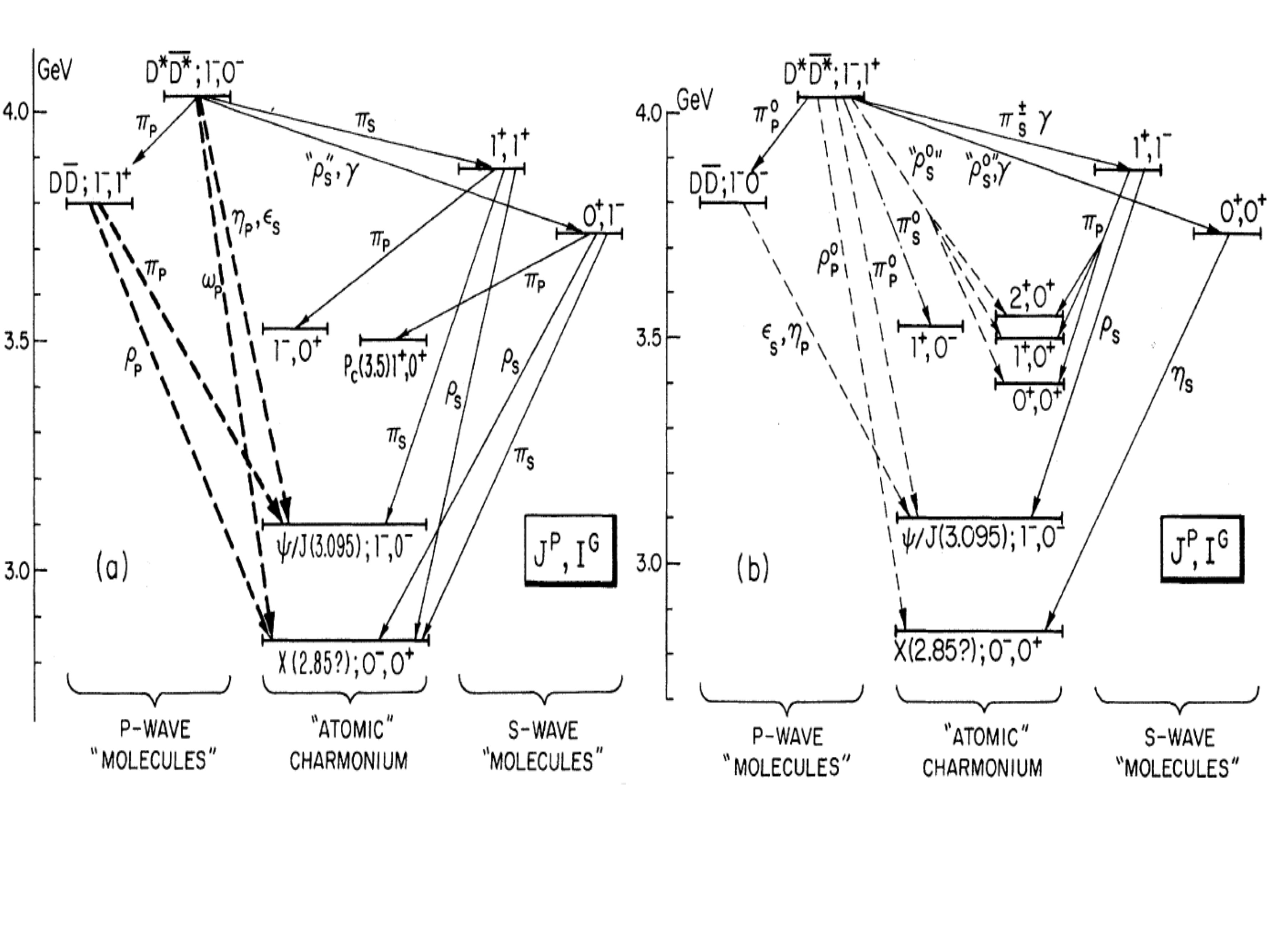}}
\caption{The spectrum of molecular charmonium\cite{OurMolecules} 
in $e^+e^-$ annihilation above 3.6 GeV.}
\label{fig-MolMess}     
\end{figure}

A currently very active endeavor is the analysis of hadrons with a larger quark
constituency than the consuetudinary old one.
The theoretical prehistory of this subject
dates back to the mid 70's \cite{Molecules1,Molecules2,Molecules3}. Perhaps the first 
data analyses with
{\it Molecular Charmonium} in mind were those of Refs.(\refcite{OurMolecules,DJ}).
In the first of these papers we concluded: {\it It seems very likely to 
us that four-quark molecules involving a $c\bar c$ pair do
exist, and have a rich spectroscopy. Our conjecture that the 4.028 GeV and
perhaps de 4.4 GeV peaks in $e^+e^-$ annihilation are indeed due to the
production of these molecules is more speculative. If it is true, then nature
has provided us with a spigot to a fascinating and otherwise almost inaccessible
new ``molecular'' spectroscopy full of experimental and theoretical challenges.}
This conclusion is still unaltered, but for  
our lack of prescience in the {\it almost inaccessible} stipulation.
That the situation would be very messy, even in a narrow energy domain 
in $e^+e^-$ annihilation, could already be concluded from Fig.~\ref{fig-MolMess}.

There have been vivid discussions on whether the observed tetra- and penta-quark resonances
are atom-like --that is, ``bags" containing all the quarks-- or molecule-like
--that is, bound states of two ``sub-bags". Studying the mesons $X$ and $Z$ 
containing two heavy quarks ($Q$) and ordinary $QQq$ baryons, Maiani, Polosa and Riquer\cite{LPR}
have recently proposed that these objects are {\it colored molecules}, respectively analogous to
the $H_2$ molecule and the $H_2^+$ ion. This is an interesting twist in the atom {\it vs} molecule
debate: a third possibility.

Amongst the scores of theoretical articles on tetra- and penta-quarks there is another one deserving in
my opinion an explicit mention. Marek Karliner and Jonathan Rosner\cite{KaRo}
predict that {\it a doubly-bottom tetraquark, $T(bb\bar u\bar d)$, with $J^P=1^+$ and $M=10389 \pm 12$ MeV
would be stable under strong and electromagnetic interactions and can only decay weakly. (It would be)
the first exotic hadron with such a property.}

\section{Back to the past again}

In the summer of '75 --after a year of upper limits incompatible with
the theoretical expectations--
evidence finally arose for the $P$-wave
 charmonia  \cite{Pc1,Pc2,Pc3}. The DESY experimentalists did
not refer to  the theorists who suggested their search; they are hereby
punished: they do not get a  reference, and they will remain
eternally ignorant of my juicy version of the story of their
competition with SLAC.

The discovery of the positronium-like $c\bar c$ spectrum of Fig.~\ref{fig-Charmonia}
started to convert many infidels to the quarker faith. And the charmed quark, not
yet found unaccompanied by its antiparticle, was
to continue playing a crucial role in the development and general
acceptance of the standard lore.
 
\subsection{R and yet another year of lank cows}

\begin{figure}[b]
\centerline{\includegraphics[width=12cm]{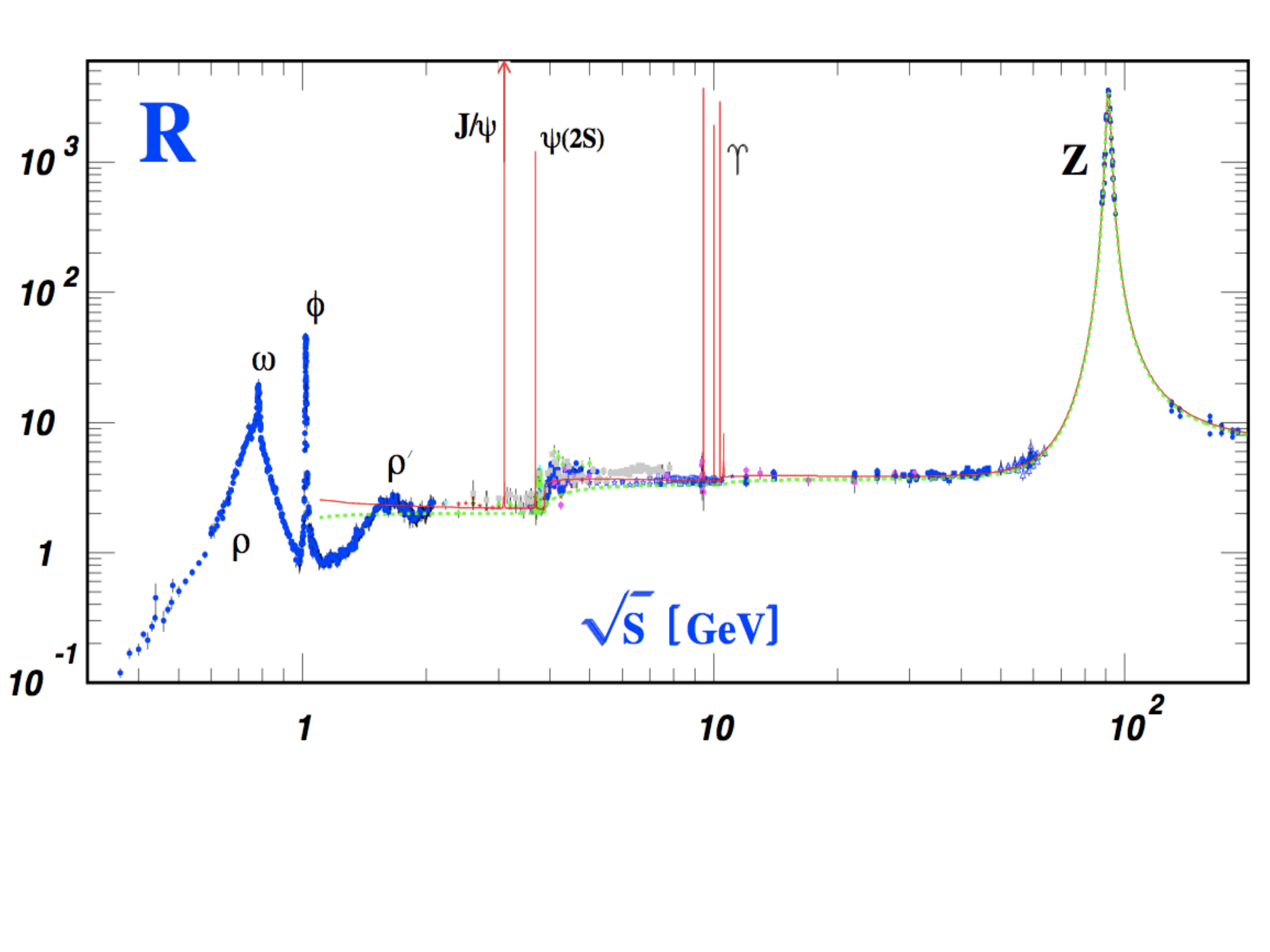}}
\vspace{-2cm}
\caption{The exquisite current data on $R(\sqrt{s})$.}
\label{fig:R}     
\end{figure}

The quantity $R\equiv \sigma(e^+e^-\to {\rm hadrons})/\sigma (e^+e^-\to  \mu^+\mu^-)$
used to be so familiar to physicists that it was unnecessary to tell the younger ones
that $R$ is not only Ricci's scalar curvature. The current observational situation for $R$ is
shown in Fig.~\ref{fig:R}, a beautiful summary of a lot of particle physics' history.
One example: between $\sqrt{s}=1$ and 3 GeV the green dotted line is the naive 
quark model prediction for $R$, with three quark flavors and the usual fractional charges.
The agreement with observation was a reason for equally naive theorists to believe in quarks.

\begin{figure}[b]
\centerline{\includegraphics[width=10cm]{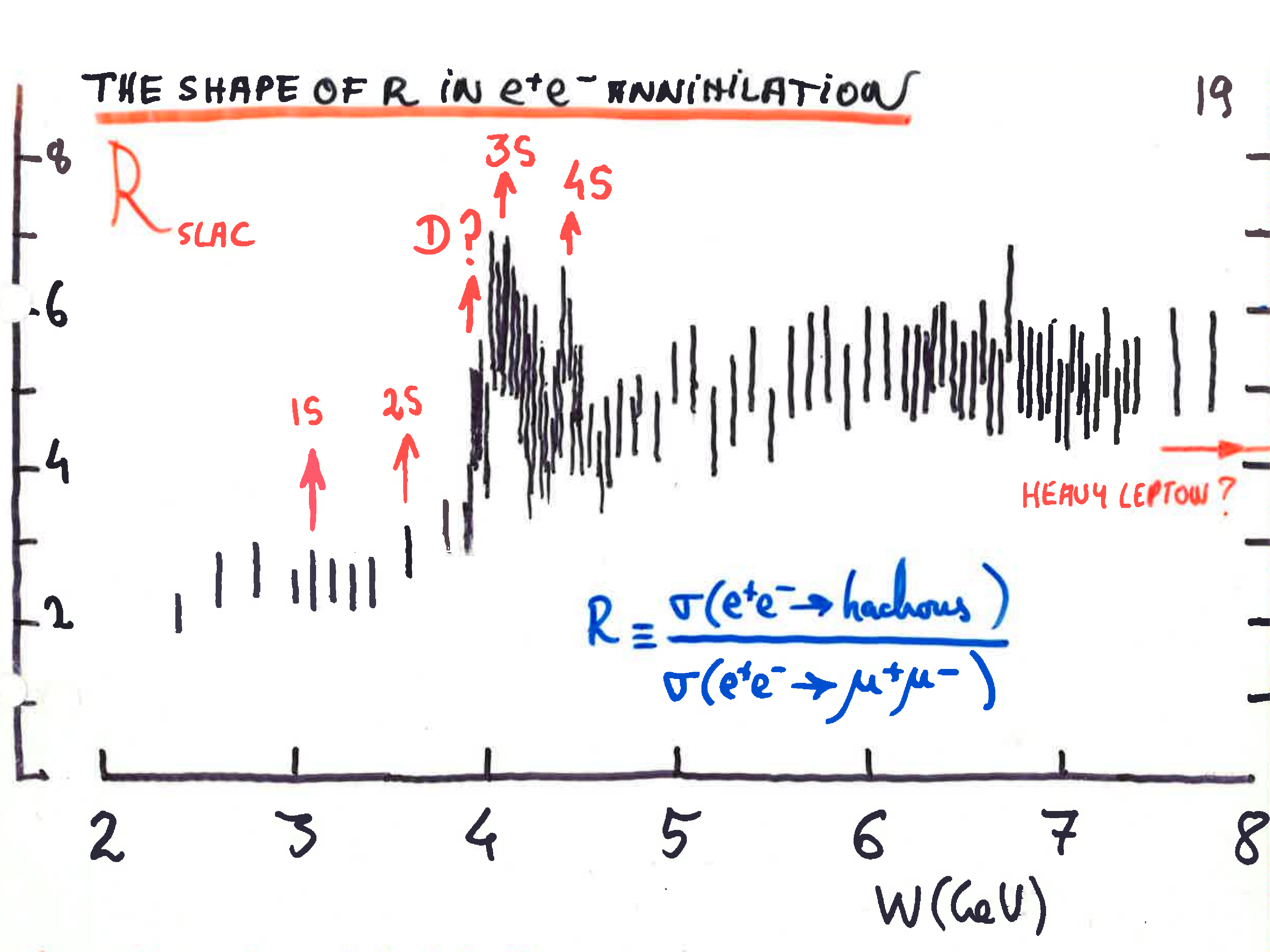}}
\caption{A vintage transparency depicting $R$ in the interesting domain.}
\label{fig-RSLAC}     
\end{figure}

In Fig.~\ref{fig-RSLAC} we see an artist's rendition of the results, in 1976,
 of measurements at SLAC of $R$. They  
showed a doubling of the yield and structure aplenty as the $\sqrt{s}\sim 4$
GeV region is crossed \cite{Sig}. Much of the jump {\bf had} to be due to the
production of charmed pairs, which were obstinately not found. Howard Georgi and I innocently
believed that a serious sharpening of the arguments would help.

In the space-like domain, $s\!<\!0$, QCD predictions 
for $e^+e^-$ annihilation are insensitive
to thresholds, bound-state singularities and hadronization caveats.
For years, theorists had been unjustifiably applying the predictions
to the time-like domain wherein experimentalists insist on taking
$e^+e^-$ data. In a paper  \cite{HYo} whose rhythmic title {\it  
Finding Fancy Flavours Counting Colored Quarks} was duly censored, we
transferred the $e^+e^-$ data, via a dispersion relation, to a
theoretically safer space-like haven.  This somersault  \cite{Adler}
allowed us to conclude that {\it the old theory with no
charm is excluded, the standard model with charm is acceptable if
heavy leptons are produced, and six quark models are viable if no
heavy leptons are produced}. Thus, anybody listening to the other
voice in the desert (that of Martin Perl, who was busy demonstrating
that he had discovered the $\tau$) had no choice but 
charm. 

Our work was improved by Enrico Poggio, Helen Quinn and Steven Weinberg  \cite{PQW}, who
realized that one could, in the complex $s$-plane, work in a contour
around the real axis where perturbative QCD can still be trusted,
whilst the distance from the dirty details of real life is judged
safe. The work of Enrico, Helen and Steve further strengthened\footnote{Or perhaps not, Helen recently
told me that their paper was not right.} our conclusion: the
measured total cross section, analyzed on firm theoretical grounds,
implied the existence of charm and of a new heavy lepton.

Imagine that some theorists, analyzing LHC data with the current Standard
Model --with its six quarks and three charged leptons-- and on the basis of a ship-shape analysis
with a statistical evidence so strong that there was no need to count $\sigma$'s,
proved that an extra quark and an extra charged lepton were being produced.
There is no doubt that the community would conclude that the cited theorists
had discovered these particles. But the social power of preconceptions cannot be overestimated.
Prior to 1976, the Standard Model --then having three established quarks and two observed
charged leptons-- was not yet accepted as ``part of the truth''. Thus, to be considered believable,
the analysis in Ref.~\refcite{HYo}  had to wait for the explicit discovery of 
open charm and the $\tau$ lepton.

\subsection{Charm is found}

No amount of published information can
compete with a few minutes of conversation. The story, whose moral
that was, is well known. For the record, I should tell it once again
 \cite{Aachen,Goldhaber,RefA}:

Shelly Glashow happened to chat with  Gerson Goldhaber in an airplane.
Surprisingly, the East Coast theorist
managed to convince the West Coast experimentalist of something.
There was no way to understand the data unless charmed particles were
being copiously produced above $\sqrt{s}=3.7$ GeV. The
experimentalists devised an improved (probabilistic) way to tell
kaons from pions. In a record 18 days two complementary SLAC/LBL
subgroups found convincing evidence for a new particle  
with all the earmarks of charm \cite{Gerson}. The charmonium
advocates at Cornell had been trying for a long time to convince the
experimentalists to attempt to discover charm by sitting on the
$\psi(3440)$ resonance, or on what would become a ``charm factory":
$\psi(3770)$ \cite{Corncharm,Corncharm2}. Alas, they initially failed.

\begin{figure}[b]
\centerline{\includegraphics[width=10cm]{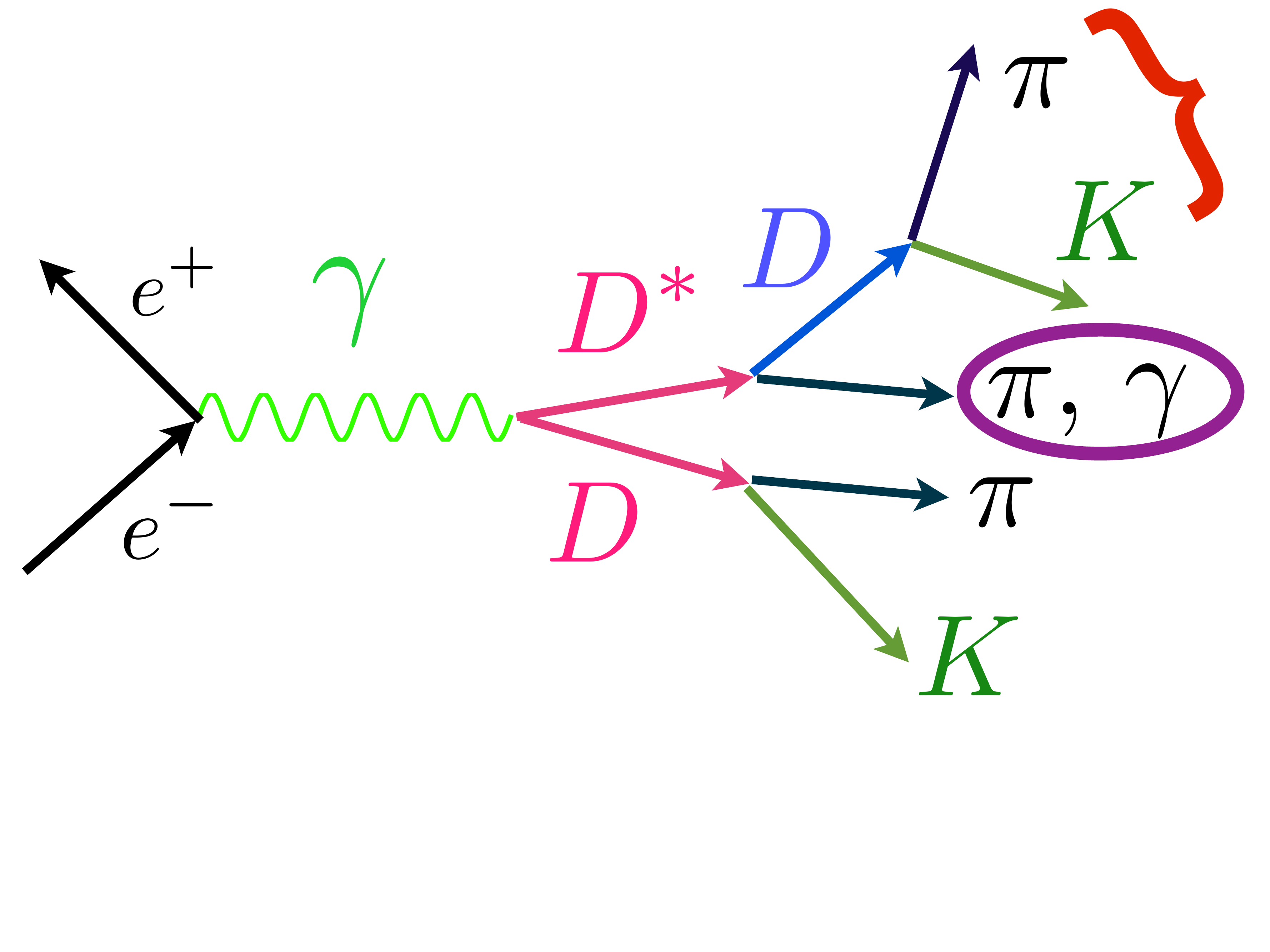}}
\vspace{-2cm}
\caption{$\bar D\, D^*$ pair-production. Invariant masses (IM)
are measured as recoiling from $K\pi$ (as in the figure), or $K\pi\pi$ ensembles.
The decay $D^*\to D\pi$ is allowed, forbidden or suppressed, depending on the particle's
charges \cite{DGG}. Close to the open-charm threshold all observed
hadrons are ``slow" and fake invariant mass peaks consequently occur.}
\label{fig-CP}     
\end{figure}

The observation of charmed mesons
ought to have been the immediate happy ending, but there was a last-minute
delay. The invariant-mass spectrum of recoiling stuff in $e^+e^-\to
D^0\,(D^\pm)+...$ had a lot of intriguing structure, but no clear
peak corresponding to $D^0\bar D^0$  or $D^+D^-$
associated production\cite{Gerson,Peru}. Enemies of the people rushed to
the conclusion that what was being found was an awful mess, and not
something as simple as charm, as in Fig.~\ref{fig-CP}.

\begin{figure}[b]
\centerline{\includegraphics[width=12cm]{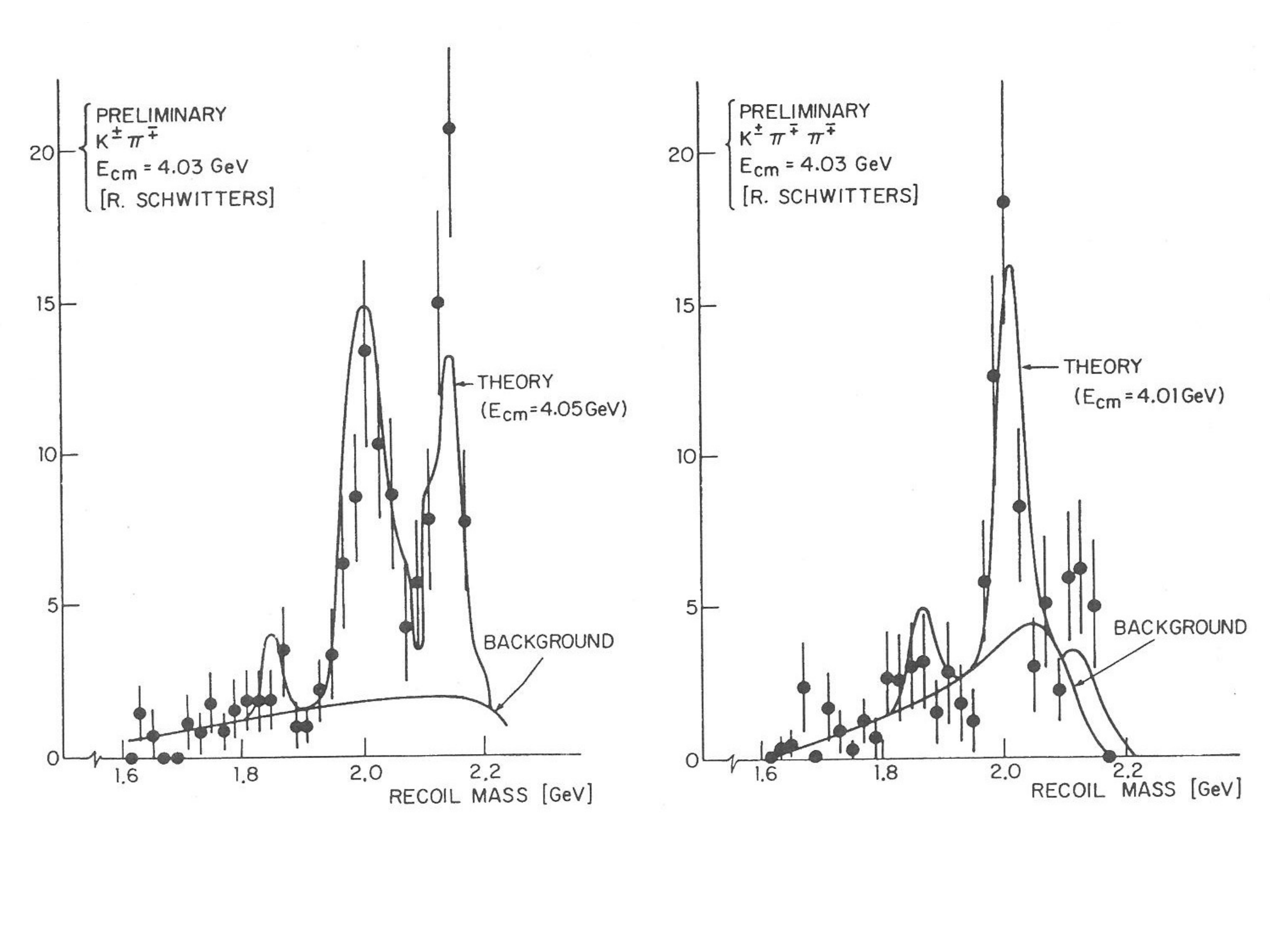}}
\caption{Predicted and observed invariant-mass spectra, recoiling
against neutral and charged $D$'s. The theoretical
curves are a one-parameter description  \cite{Charm}.}
\label{fig-CD}     
\end{figure}

But we had one  last unspent cartridge  \cite{Charm}. We expected $D\bar
D$, $D\bar D^*+\bar D D^*$, and $D^*\bar D^*$ production to occur in
the ``spin'' ratio 1:4:7 (thus the $D\bar D$ suppression). We trusted
our prediction  \cite{DGG} $m(D^*)-m(D)\simeq m(\pi)$, which implies
that for charm production close to threshold, the decay pions are
slow and may be associated with the ``wrong'' $D$ or $D^*$ to produce
fake peaks in recoiling mass. Finally, we knew that the charged $D$'s
and $D^*$'s ought to be a little heavier than their neutral sisters. Consequently, the
$D^*$ decays had to be very peculiar: $D^{*0}\to D^+\pi^-$ is
forbidden, $D^{*0}\to D^0\gamma$ competes with $D^{*0}\to D^0\pi^0$,
etc. On the basis of these considerations (and with only one fit
parameter) we constructed the recoil spectra shown in Fig.~\ref{fig-CD}.
 Case closed!

\subsection{Seeing (gluons) is believing}

Quarks have not been seen and may  \cite{DGJ} never be. But their 
manifestation as quark jets was apparent since 1976  \cite{Qjets}.
Kogut and Susskind  \cite{KS} argued that the gluon 
could show up in
the same way: the elementary process $e^+e^-\to \bar q\, q\, g$ may
result in three-jet final states.

Further work on QCD jets was often based on ``intuitive perturbation
theory'', an appellation perhaps meant to admit a fundamental
lack of understanding. Decorum was regained by
the work of Sterman and Weinberg  \cite{SW}, Georgi and Machacek
 \cite{HoMa} and Farhi  \cite{EF}, who exploited the fact that in QCD,
as in QED, there are ``infrared-safe'' predictions  \cite{HDP}, not
sensitive to the long-distance dynamics that, in QCD, are intractable
in perturbation theory.

\begin{figure}[]
\centerline{\includegraphics[width=9cm]{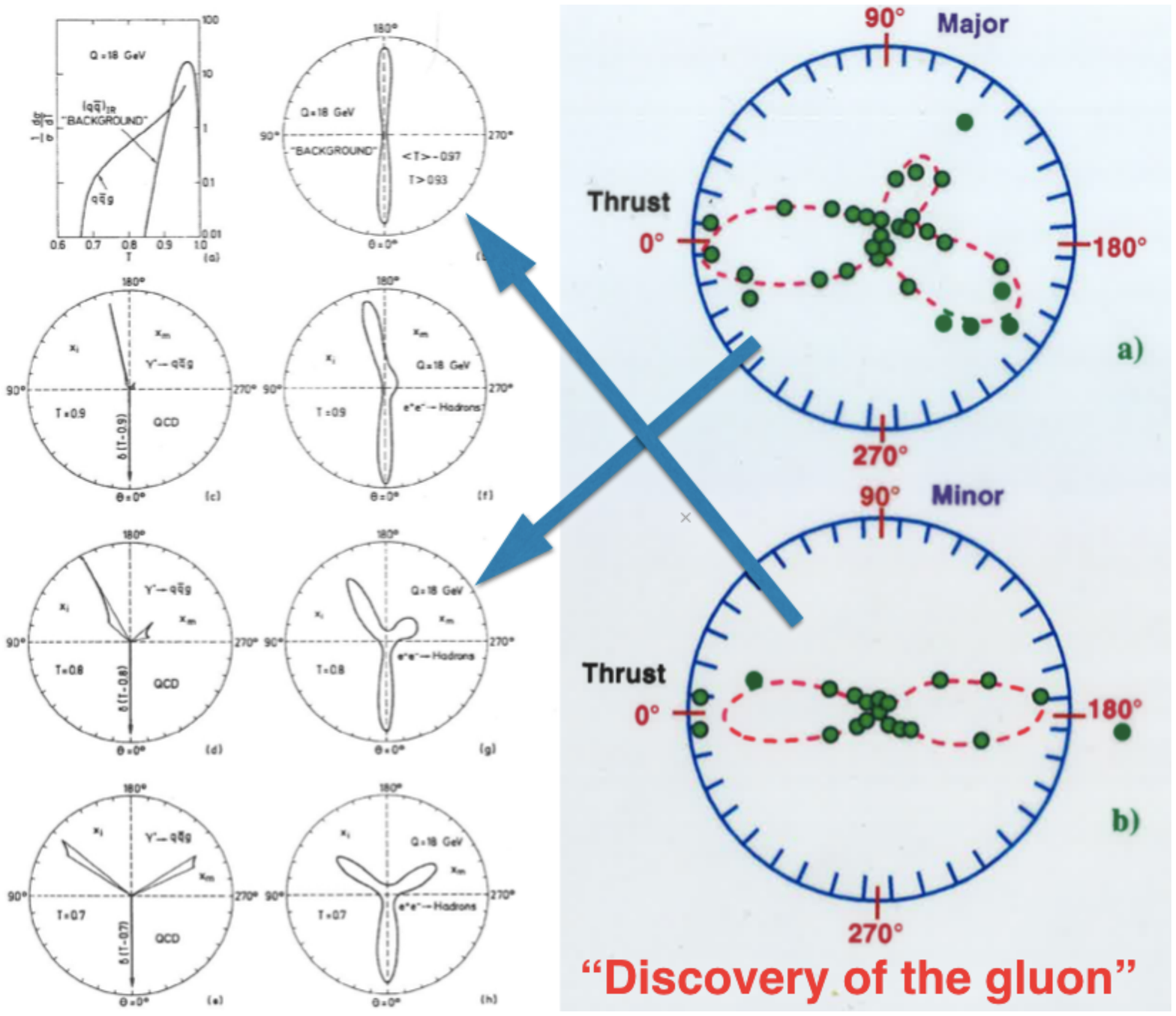}}
\vspace*{-4cm}
\caption{Examples of predictions \cite{DEFG}  for three-jet
distributions at different fixed thrusts, $T$. The left column contains
the leading perturbative QCD predictions; in the middle one the
results are smoothed for ``hadronization" effects. More often than not,
the small jet is produced by a gluon. The right panel contains results from the 
Mark-J Collaboration\cite{MarkJ}.}
\label{Gluon}
\end{figure}

One infrared-safe observable is the ``antenna'' pattern of energy
flow in an ensemble of hadronic final states in $e^+e^-$
annihilation, properly reoriented event by event to compensate for
the vicious quantum mechanical penchant for uncertainty.  We foretold
 \cite{DEFG} the pattern, binned in ``thrust''  \cite{EF},  to be that
of Fig.~\ref{Gluon}. This three-jet structure and the QCD-predicted details of
the angular or energy distributions played an important role in the
``discovery of the gluon'', a subject whose denouement 
(that gluons are for real) has been
described in detail by James Branson in Ref.~\refcite{JB}.

\subsection{The triple gluon coupling}

\begin{figure}[]
\centerline{\includegraphics[width=8cm]{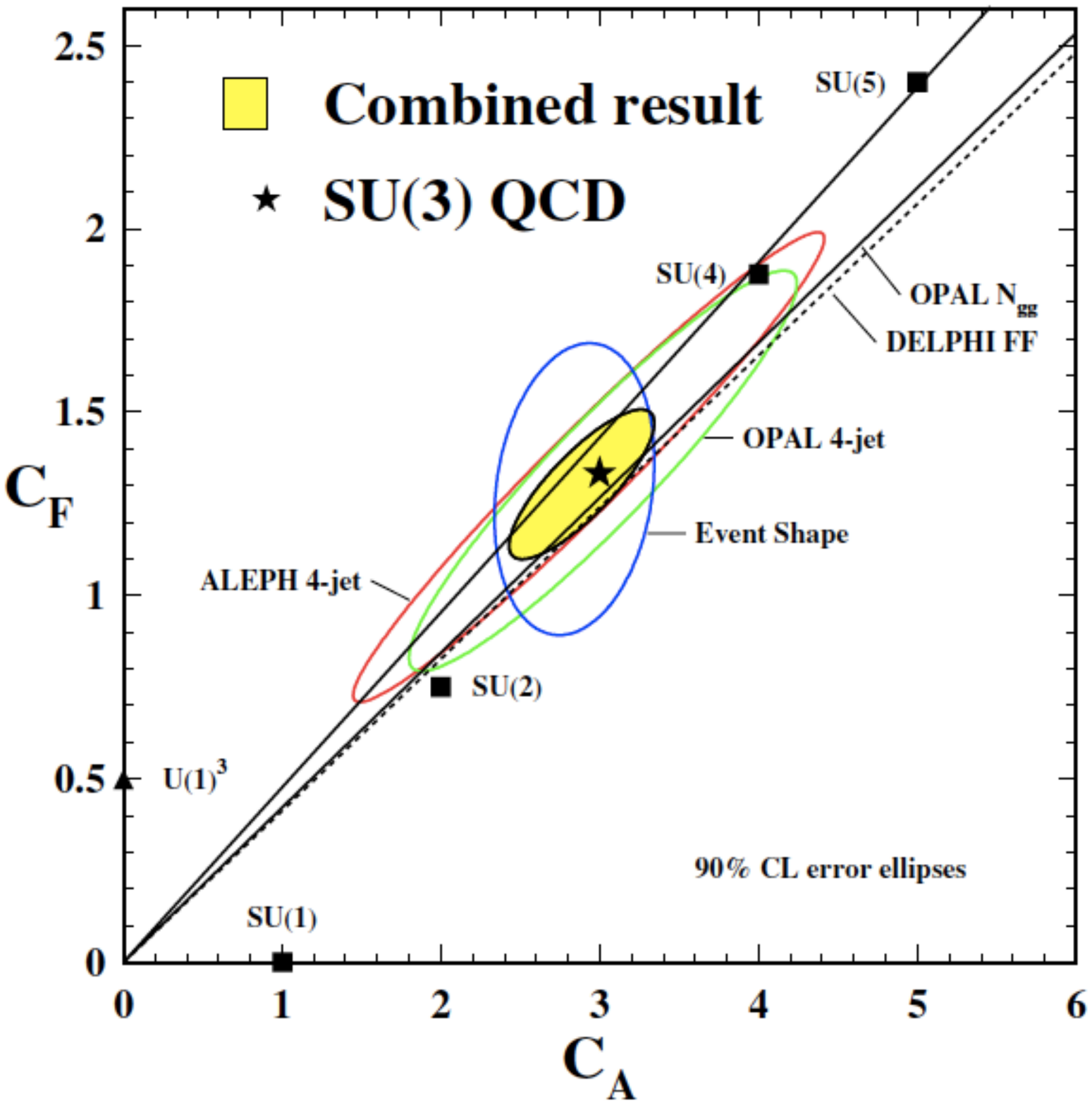}}
\vspace*{-2cm}
\caption{Measurements of the color ``Casimirs" of quarks and gluons\cite{Casimirs}.}
\label{fig:Casimirs}
\end{figure}

Once upon a time, a measurement of the triple gluon coupling was thought to be nearly 
impossible\cite{YoPL}. But that was before the advent of LEP. Recall that $C_A=N_c=3$
is the triple-gluon ``Casimir" color factor associated with gluon emission
by a gluon, $C_F=(N_c^2-1)=4/3$  is the one for gluon emission by a quark and
$T_R=1/2$ is for a gluon to split into $q\bar q$. Fixing $T_R$, $\alpha_s$ and the
number of flavors, it is possible\cite{Casimirs} to extract $C_A$ and $C_F$ from
the analysis of event shapes and other observables
at LEP. The result of combining all experiments as
in Fig.~\ref{fig:Casimirs} is 
$C_A= 2.89\pm 0.03\,\rm (stat)\pm 0.21\,(syst)$; $C_F= 1.30\pm0.01\,\rm(stat)\pm 0.09\,(syst)$.

The above results, impressive as they are, do not compete in precision and cleanliness
 with the corresponding measurement in General Relativity, discussed in Section \ref{sec:GR}.

\section{Confronting reality in full detail}

\begin{figure}[]
\centerline{\includegraphics[width=10cm]{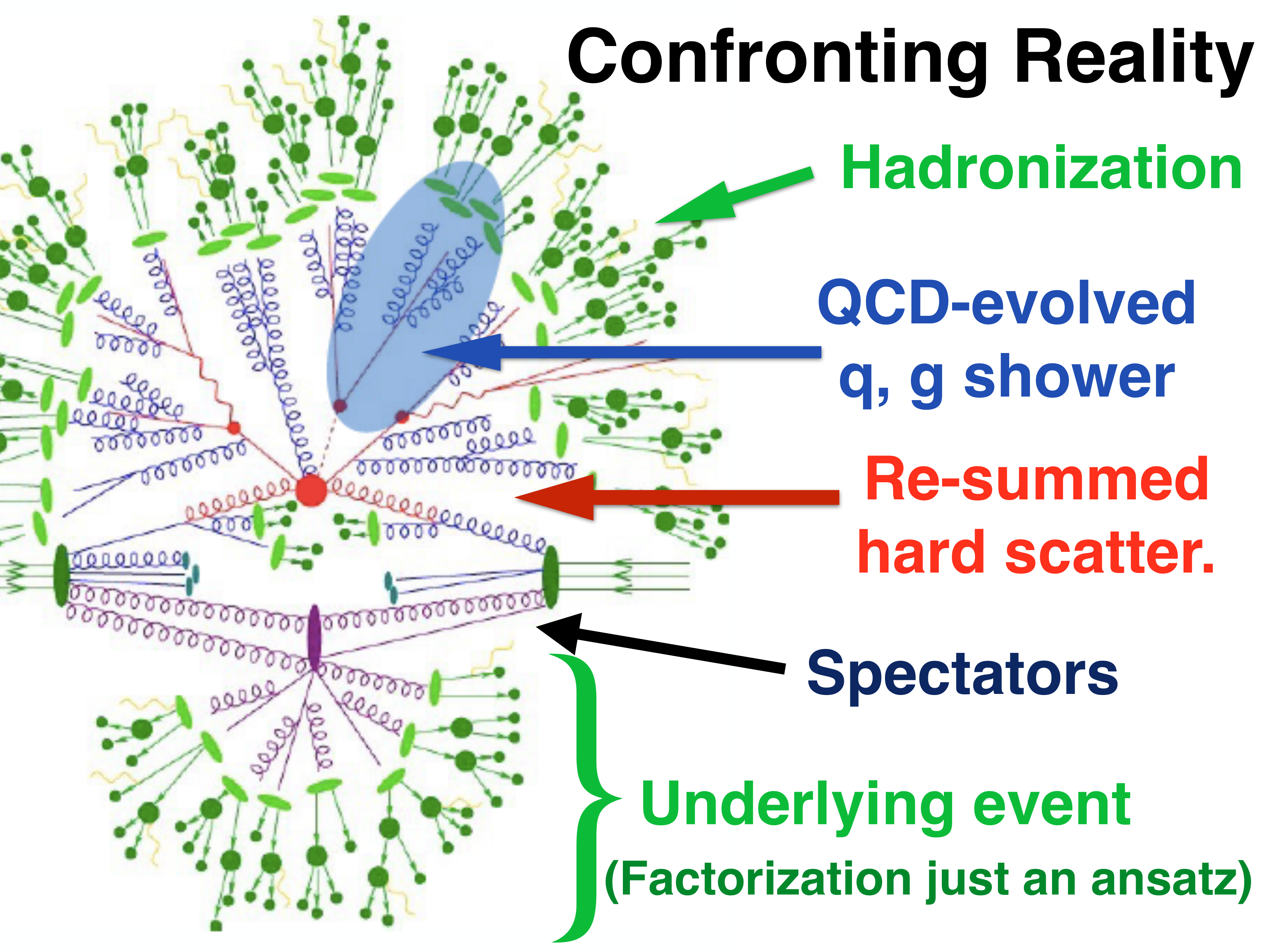}}
\caption{The hardships involved in facing reality in scrupulous detail.}
\label{fig:CR}
\end{figure}

I have not been able to find the originator of the intimidating drawing in Fig.~\ref{fig:CR}.
Its green parts (hadronization and the ``underlying" event) are not understood at the quite satisfactory
level of its hard scattering and QCD shower parts. They are treated by ``phenomenological"
methods requiring a back-and-forth zitterbewegung between predictions and observations, to tune
parameters and/or coach a machine learning program. The hadronization issue is tackled by
non-quantum-mechanical methods, such as stretching frangible strings or gathering colorless clusters.
For a review of this immense subject, see Ref.~\refcite{NaSk}.

\section{Themes left uncovered}

Amongst others: the QCD phase diagram, chiral dynamics and symmetry breaking,
heavy-quark methods 
and the so-called ``Early Universe Plasma", with the quotes referring to the fact that the
QCD plasma studied in $pp$ collisions has, locally, boundaries that the Universe did not have
(which is a subject of concern and discussion) and does not have a thermally 
equilibrated bath of neutrinos (which is quite irrelevant).
 
The only remaining particle predicted by the Standard Model and not yet found
is the axion. And this is definitely not because of a lack of attempts to produce
and detect it in the lab, or to observe it as a plausible constituent of dark matter.
 
The most challenging QCD problem --confinement-- is not yet solved,
in spite of a one M\$ prize awaiting whoever solves it \cite{Clay}.
Almost all we have concerning confinement is the good old  ``stringy" lattice result for the potential
$V(r)$ between two static quarks, $V(r)\to \sigma\, r+{\rm const}/r$, as $r\to \infty$,
with $\sigma$ the string's tension.

\section{One moral of the story}

In the olden days experimentalists, particularly the ones working in the 
West Coast of the USA, were strongly motivated to disprove all
theories and to mistrust almost all theorists; they were
perhaps permeated by some arcane Californian
faith that nature is intrinsically unfathomable. They definitely
did not have in their data-analysis programs the current instruction
stating: {\it iff [result = standard; look ``elsewhere"]}. This made life
most enjoyable and the case for the once-upon-a-time generally ignored Standard Model extremely 
strong. To gauge whether or not things have turned around, consider supersymmetry.

At various points in this personalized rendering of the history of QCD I have grumbled
about experimentalists not citing the ``phenomenological" papers that made their searches
and even their findings possible. This continues to be very much the case. One example:
for years on end, ``QCD phenomenologists"
have made an enormous collective effort  to provide predictions 
beyond the tree level ones for a host of processes. More often than not, their work is referred to
by experimentalists as the ``Standard Model (SM) prediction". But the SM does {\bf not}
make predictions. Specific people use the SM to make predictions. A hypothetical reader who has
reached this far is presumably a QCD phenomenologist. I have no doubt
that (s)he would agree with me on all this.

A recent example of disregard of crucial phenomenological inputs by experimentalists
has to do with the discovery of the Higgs boson. A question to ask is: why was it announced as a
candidate for such a particle? The first and foremost answer is that its production cross section
agreed with the expected one. But the dominant production mechanism is a highly non-trivial
gluon fusion via a heavy ``triangular" quantum loop\cite{FWTriangle}. 
Georgi, Glashow, Maria Machacek and Dimitri
Nanopoulos\cite{GGMN}, who first calculated the $pp\to H+...$ cross section via gluon fusion, 
were not cited in either of the Higgs
discovery talks. 

A second point is that
one of the two discovery channels was $H\to\gamma\gamma$. It is also a non-trivial one-loop
quantum effect. The corresponding decay width was first computed by John Ellis, Mary K.
Gaillard and Nanopolous\cite{JMKN}. Although in this particular case the experimentalists may have had
an excuse not to quote these authors\footnote{The closing sentence in Ref.~(\refcite{JMKN}) is:
{\it ... we do not want to encourage big experimental searches for the Higgs boson, but we do feel
that people performing experiments vulnerable to the Higgs boson should know how it may
turn up.}} the fact is that they did not.

Finally, there is a novel recently discovered way of not quoting the authors of a pivotal contribution.
Close to the arbitrary $5\,\sigma$ ``discovery level" several things obviously play a role: the
detectors' characteristics, pure statistical luck or lack of luck, and the analysis methods. So?

The second Higgs-discovery channel was $H\to \ell_1^+\,\ell_1^-\,\ell_2^+\,\ell_2^-$, with $\ell=e$ or $\mu$.
There are seven relevant independent observables in this process,
beautifully entangled in their 7D space, but differently so for the signal and the background.
Accumulated event by event, the likelihood ratio of signal to background is the optimal tool
(now called ``the matrix element method") to enhance the significance of a result. For a Higgs boson
(or other objects with various $J^{PC}$ values)
lighter than two ``on-shell" $Z$s the methodology and relevant results were procured by the authors of 
Ref.~\refcite{DLPRS}. I am not quoting their names in text because that is what Joe Incandela
did in his CMS Higgs discovery talk. But he showed the Physical Review
 reference without names, as if authorship
went without saying, like in the case of Hamlet or La Gioconda. In his next transparency Joe showed
that the significance contributed by the $H\to 4\,\ell$ channel was crucial in their having (barely) reached
$5\,\sigma$. 

It is difficult not to feel that particle physics
phenomenology is generally less appreciated than it should,  particularly
in comparison with theoretical work ``beyond'' this or that. 

\section{Windup}

There has been an enormous progress in perturbative and non-perturbative QCD
since quarks were invented, in late 1963.
This progress often required phenomenally difficult
theoretical or phenomenological developments, and the later played a key role 
in the understanding of experimental results, most recently at the LHC. 

Theories with gauge degrees of freedom have played the central role in our understanding
of nature for a very long time. Interestingly, almost all of the most challenging currently
identified problems pertain to these theories: confinement, the ``naturalness" of the
Higgs boson mass and the question of renormalizability, in QCD, the Electroweak Theory
and General Relativity.

\vspace{.6cm}

\noindent {\bf Acknowledgments.} 
I am most indebted to my collaborators, in particular the ones I have
cited. 
This project has received funding/support from the European Union's Horizon 2020 research
and innovation program under the Marie Sk\l idowska grant agreement No 674996.

%
%
%

\end{document}